\documentclass[twocolumn]{aastex7}

\begin{document}

\title{Which active galaxies might be neutrino emitters?}

\correspondingauthor{Mouyuan Sun}
\email{msun88@xmu.edu.cn}

\author[orcid=0009-0005-2801-6594]{Shuying Zhou}
\affiliation{Department of Astronomy, Xiamen University, Xiamen, Fujian 361005, People’s Republic of China; msun88@xmu.edu.cn}
\email{zhoushuying@stu.xmu.edu.cn}

\author[orcid=0000-0002-0771-2153]{Mouyuan Sun} 
\affiliation{Department of Astronomy, Xiamen University, Xiamen, Fujian 361005, People’s Republic of China; msun88@xmu.edu.cn}
\email{msun88@xmu.edu.cn}  

\author[0000-0002-0092-7944]{Guobin Mou}
\affiliation{Department of Physics and Institute of Theoretical Physics, Nanjing Normal University, Nanjing 210023, People’s Republic of China}
\email{gbmou@njnu.edu.cn}

\author[0000-0003-1474-293X]{Da-bin Lin}
\affiliation{Guangxi Key Laboratory for Relativistic Astrophysics, School of Physical Science and Technology, Guangxi University, Nanning 530004, People’s Republic of China}
\email{lindabin@gxu.edu.cn}

\author[orcid=0000-0001-8678-6291]{Tong Liu} 
\affiliation{Department of Astronomy, Xiamen University, Xiamen, Fujian 361005, People’s Republic of China; msun88@xmu.edu.cn}
\email{tongliu@xmu.edu.cn}

\author[0000-0002-9067-3828]{Ming-Xuan Lu} 
\affiliation{Guangxi Key Laboratory for Relativistic Astrophysics, School of Physical Science and Technology, Guangxi University, Nanning 530004, People’s Republic of China}
\email{lumx@st.gxu.edu.cn}

\author[orcid=0000-0002-1935-8104]{Yongquan Xue} 
\affiliation{Department of Astronomy, University of Science and Technology of China, Hefei 230026, People’s Republic of China}
\affiliation{School of Astronomy and Space Science, University of Science and Technology of China, Hefei 230026, People’s Republic of China}
\email{xuey@ustc.edu.cn}

\begin{abstract}
The IceCube Neutrino Observatory has identified several individual neutrino emitters associated with supermassive black hole accretion phenomena, including blazars, tidal disruption events, and, unexpectedly, Seyfert galaxies. A key open question is which types of active galactic nuclei (AGNs) are most likely to be neutrino emitters. Here we show that high-confidence extragalactic neutrino emitters tend not only to have higher hard X-ray fluxes but also to be more variable in mid-infrared (MIR) than other AGNs in the \textit{Swift} BAT AGN Spectroscopic Survey. MIR variations effectively trace long-term fluctuations in AGN accretion disks and/or jets. In addition to the role of X-ray flux emphasized in previous studies, we speculate that long-term central engine fluctuations may also be critical for neutrino production. This hypothesis may inform IceCube neutrino-electromagnetic counterpart association studies and provide new insights into cosmic ray acceleration sites. First, the observed neutrinos are unlikely to originate from AGN host galaxies or from interactions between large-scale (dozens of parsecs) winds/outflows and the surrounding interstellar medium. Second, if neutrinos are produced in the X-ray corona, the corona should exhibit strong magnetic turbulence dissipation or magnetic reconnection whose rate changes substantially on timescales of years. Third, the relativistic jets of blazar neutrino emitters may be intrinsically unstable over years. Finally, if neutrinos are related to interactions between small-scale winds/outflows and torus clouds, such winds/outflows must be highly episodic. 
\end{abstract}


\section{Introduction} \label{sec:intro}
Neutrinos, interacting weakly with particles or magnetic fields, provide a powerful probe of extreme environments. After detecting the diffuse high-energy astrophysical neutrinos \citep{IceCube2013}, the IceCube Neutrino Observatory has recently identified several electromagnetic (EM) counterparts to extragalactic high-energy (TeV--PeV) neutrinos \citep[e.g.,][]{IceCube2018a-TXS, IceCube2022-1068, Abbasi2025-1068-4151}. Such neutrinos are thought to originate from high-energy cosmic rays via lepto-hadronic ($p\gamma$) and purely hadronic ($pp$) interactions. Active galactic nuclei (AGNs), long regarded as promising cosmic-ray accelerators, are natural candidates for neutrino emitters \citep[e.g.,][]{Begelman1990, Stecker1991, Halzen1997}. 

The EM counterparts of IceCube neutrinos have been associated with different types of AGNs. The first extragalactic high-energy neutrino emitter identified by spatial and temporal coincidence is TXS 0506+056, a gamma-ray flaring blazar \citep{IceCube2018a-TXS, IceCube2018b-TXS}. Relativistic jets in blazars that are oriented close to our line of sight are promising sites for cosmic ray acceleration \citep[e.g.,][]{Ansoldi2018-blazar-jet, Keivani2018-blazar-jet, Murase2018-blazar-jet, Cerruti2019-blazar-jet, Gao2019-blazar-jet}. With the discovery of jetted tidal disruption events \citep[TDEs; e.g.,][]{Burrows2011, Zauderer2011}, by analogy with blazars, TDEs are recognized as potential counterparts for high-energy neutrinos \citep[e.g.,][]{Wang2011, Lunardini2017, Liu2020-TDE-jet, Winter2021-TDE-jet}. Several TDEs show potential neutrino associations via spatial and temporal correlations \citep[e.g.,][though \citealt{Zegarelli2025} note potential mismatches]{Stein2021-TDE-dsg, Reusch2022-TDE-fdr, Jiang2023, van_Velzen2024-TDE-aalc, Yuan2024-AT2021lwx}. 

Interestingly and unexpectedly, the Seyfert galaxy NGC 1068 has been identified as a high-confidence extragalactic point source of neutrino emission \citep{IceCube2022-1068, Abbasi2025-1068-4151}. In addition, stacking analyses indicate that the high-energy neutrino emission from blazars is insignificant compared to that from non-blazars \citep{Abbasi2025-1068-4151}. Seyferts are typically radio-quiet AGNs that possibly have weak jets. Hence, it is entirely unexpected that Seyfert galaxies are efficient in accelerating cosmic rays and emitting neutrinos. 

Although only a few Seyferts have been claimed as neutrino emitters, several mechanisms have been proposed to explain neutrino production in Seyferts. Candidate sites include the hot corona \citep[e.g.,][]{Inoue2019-AGN-corona, Inoue2020-AGN-corona, Murase2020-AGN-corona, Kheirandish2021-AGN-corona, Murase2022-AGN-corona}, radiatively inefficient accretion flows \citep[RIAFs; see, e.g.,][]{Kimura2019-AGN-IRAF, Kimura2021-AGN-IRAF, Gutierrez2021-AGN-IRAF}, interactions between AGN outflows/winds and surrounding materials \citep[e.g.,][]{Inoue2022-AGN-interact, Fang2023-AGN-interact, Huang2024-AGN-interact, Ehlert2025-AGN-interact}, and embedded stars or compact objects within the AGN accretion disk \citep[e.g.,][]{Zhu2021a-AGN-AMS, Tagawa2023-AGN-AMS}. These processes, in addition to relativistic jets, have also been suggested as possible neutrino production channels in blazars \citep{Kun2024-AGN-blazar, Zathul2024-AGN-blazar, Yang2025-AGN-blazar, Fiorillo2025-AGN-blazar} and in TDEs \citep{Hayasaki2019-TDE-accretion, Fang2020-TDE-interact, Murase2020-TDE-corona, Mou2021, Wu2022-TDE-interact, Winter2023-TDE-interact}. 

Naturally, an important question arises: \textit{Which AGNs are most likely to emit neutrinos?} Addressing this question is essential both (1) for guiding efficient EM counterpart identification via stacking techniques given weak signals, and (2) for testing the proposed production mechanisms. For example, the hot corona scenario \citep[e.g.,][]{Inoue2020-AGN-corona, Murase2020-AGN-corona} predicts a correlation between X-ray flux and neutrino flux \citep[e.g.,][]{Kun2024-AGN-blazar, Fiorillo2025-AGN-corona}. This prediction has motivated searches for neutrino sources within hard X-ray AGN catalogs \citep{Neronov2024-AGN-3079, Abbasi2024-420-015, Abbasi2025-1068-4151}. It also offers a natural explanation for the neutrino association with NGC 1068, whose absorption-corrected X-ray flux is higher than that of most AGNs. However, several AGNs exhibit intrinsic X-ray fluxes exceeding that of NGC 1068 (see Section~\ref{sec:bass}), yet no neutrinos have been detected from them. \textit{Therefore, why NGC 1068 is currently the brightest known extragalactic neutrino source remains unresolved.} 

Motivated by the possibility of efficient cosmic-ray acceleration in strong shocks or violent magnetic dissipation, we speculate that AGNs with strong long-term variability and high X-ray fluxes are promising candidates for neutrino emission. The time-integrated identification of high-energy neutrino emitters often relies on decade-long IceCube monitoring \citep[e.g.,][]{Abbasi2025-1068-4151, Abbasi2024-420-015}; hence, it should be sensitive to neutrino production mechanisms that operate on long timescales of years. Mid-infrared (MIR) variability, which naturally probes long-term central engine fluctuations, therefore serves as a suitable proxy for investigating the possible AGN-IceCube neutrino association. 

The X-ray/UV/optical emission from the AGN accretion disk can be absorbed by the dusty torus and re-emitted in the MIR band. For non-jetted AGNs, the dust responsible for MIR emission typically resides at parsec scales from the central supermassive black hole \citep[e.g.,][]{Jaffe2004, Tristram2009}. As a result, MIR light curves do not reflect the central engine instantaneously; instead, they respond with time lags of $\sim 10$--$10^3$ days \citep[e.g.,][]{Lyu2019, Mandal2024} and exhibit smoothed patterns relative to the X-ray/UV/optical light curves, with short-term variability strongly suppressed \citep[e.g.,][]{Li2023}. Strong host galaxy starlight contamination can further dilute the observed variability amplitudes. Nevertheless, MIR emission still preserves the long-term variability of the central engine and is hardly affected by dust extinction. In jetted AGNs, the observed MIR emission may be a mixture of torus and jet; thus, the MIR variations may also trace jet fluctuations in blazars. In summary, MIR variability provides an effective tracer of long-term fluctuations in AGN central engines or relativistic jets, regardless of dust obscuration. 

In this manuscript, we use AGNs in the second data release of the \textit{Swift} BAT AGN Spectroscopic Survey \citep[BASS DR2;][]{Koss2022} to show that AGNs with neutrino associations indeed not only have large intrinsic X-ray fluxes but also display long-term variability in the MIR. The manuscript is structured as follows. Section~\ref{sec:data} introduces the sample, Section~\ref{sec:res} presents the results, and Section~\ref{sec:discus} discusses the implications. In Section~\ref{sec:conclusion}, we summarize the main conclusions.

\section{Data} \label{sec:data}
\subsection{Neutrino sources}
\subsubsection{Blazars}
TXS 0506+056 is the first blazar identified with a significance greater than $3\sigma$ in association with high-energy neutrinos \citep{IceCube2018a-TXS, IceCube2018b-TXS}. Subsequently, \citet{Aartsen2020-10yr} perform a time integrated search using $10$ years of IceCube data for $110$ sources and report a $3.3\sigma$ significance in the northern hemisphere, associated with the Seyfert galaxy NGC 1068 and three blazars, TXS 0506+056, PKS 1424+240, and GB6 J1542+6129. Using time-dependent binomial tests on the same IceCube dataset and AGN sample, \citet{Abbasi2021-10yr2} also report a neutrino excess in the northern hemisphere at the $3\sigma$ level, associated with M87, NGC 1068, TXS 0506+056, and GB6 J1542+6129. Three of these sources overlap with the time-integrated results of \citet{Aartsen2020-10yr}. For our analysis, we focus on the two blazars identified in both the time-integrated analysis and the time-dependent binomial tests, i.e., TXS 0506+056 and GB6 J1542+6129. 

The hard X-ray fluxes of TXS 0506+056 and GB6 J1542+6129 are estimated from the $15-55\ \mathrm{keV}$ fluxes reported by \citet{Kun2024-AGN-blazar} based on \textit{NuSTAR} observations. For TXS 0506+056, the $15-55\ \mathrm{keV}$ flux is $2.39\times 10^{-12}\ \mathrm{erg\ s^{-1}\ cm^{-2}}$ with a photon index of $1.5$--$1.9$ (we adopt $1.7$). For GB6 J1542+6129, the $15-55\ \mathrm{keV}$ flux and photon index are $6.0\times 10^{-13}\ \mathrm{erg\ s^{-1}\ cm^{-2}}$ and $1.55$, respectively \citep{Kun2024-AGN-blazar}. By extrapolating the $15-55\ \mathrm{keV}$ fluxes to the $14-150\ \mathrm{keV}$ band via the power-law spectra, we obtain $14-150\ \mathrm{keV}$ fluxes ($f_{14-150\ \mathrm{keV}}$) of $5.09\times 10^{-12}\ \mathrm{erg\ s^{-1}\ cm^{-2}}$ for TXS 0506+056 and $1.40\times 10^{-12}\ \mathrm{erg\ s^{-1}\ cm^{-2}}$ for GB6 J1542+6129. We note that these hard X-ray fluxes have not been corrected for (presumably weak) absorption by the host galaxy. 

\subsubsection{Seyfert galaxies}\label{sec:seyfert}
The Seyfert galaxy NGC 1068 has been identified as a high-confidence TeV neutrino source \citep{IceCube2022-1068}. Subsequently, \citet{Neronov2024-AGN-3079} report that both NGC 4151 and NGC 3079 may be neutrino sources, while \citet{Abbasi2024-420-015} find an excess of neutrinos from NGC 4151 and CGCG 420-015. In addition, \citet{Abbasi2025-1068-4151} also report an excess associated with NGC 4151. However, the evidence for NGC 3079 is of relatively lower confidence \citep{Abbasi2025-1068-4151}, and the significance of CGCG 420-015 depends strongly on the adopted spectral model \citep{Abbasi2024-420-015}. All of these searches consistently report NGC 4151 as a neutrino source. A potential caveat, however, is raised by \citet{Omeliukh2025}, who note the presence of two blazars close to NGC 4151 that could contribute to the observed neutrinos. Detailed modeling shows that neutrinos from these two blazars would peak above $\sim 100$ PeV, making their contribution to the TeV neutrinos observed from NGC 4151 negligible \citep{Omeliukh2025}. Thus, we speculate that NGC 4151 remains a confident neutrino source.

The intrinsic hard X-ray fluxes ($f_{14-150\ \mathrm{keV}}$) for NGC 1068 and NGC 4151 are taken from the $70$-month \textit{Swift}/BAT survey \citep{Ricci2017, Koss2022}, with values of $1.79\times 10^{-10}\ \mathrm{erg\ s^{-1}\ cm^{-2}}$ and $4.59\times 10^{-10}\ \mathrm{erg\ s^{-1}\ cm^{-2}}$, respectively. Notably, \cite{Inoue2020-AGN-corona} argue that the intrinsic X-ray flux of NGC 1068 exceeds that of NGC 4151 by a factor of $3.6$, in contrast with the values adopted here. This discrepancy is probably due to NGC 1068 being a highly obscured AGN, and its absorption-corrected X-ray flux is difficult to reliably recover. \textit{For consistency, the intrinsic fluxes for NGC 1068 and NGC 4151 that we adopt are from the same team \citep{Ricci2017} and are based on the same 70-month Swift/BAT survey.}

\subsection{The BASS DR2 sample}\label{sec:bass}
The absence of significant gamma-ray emission associated with the neutrino fluxes in NGC 1068 \citep{Acciari2019} and NGC 4151 \citep{Murase2024-4151} has motivated theories suggesting that neutrinos are produced in the AGN corona, where gamma rays above $100\ \mathrm{MeV}$ are attenuated through the $\gamma\gamma \to e^+e^-$ process \citep[e.g.,][]{Inoue2019-AGN-corona, Inoue2020-AGN-corona}. The coronal origin scenario, in turn, motivates searches for neutrino sources in hard X-ray AGN catalog \citep{Neronov2024-AGN-3079, Abbasi2024-420-015, Abbasi2025-1068-4151}. We therefore compare the intrinsic hard X-ray fluxes ($f_{14-150\ \mathrm{keV}}$) of AGN neutrino emitters with other AGNs. We adopt the hard X-ray selected AGNs from the BASS DR2 catalog \citep{Koss2022}, whose intrinsic X-ray fluxes are measured \textit{consistently} by \cite{Ricci2017}. 

The BASS DR2 catalog contains $858$ AGNs. Following previous work \citep{Neronov2024-AGN-3079}, we restrict our analysis to the $409$ AGNs with declinations between $-5^\circ$ and $60^\circ$, a range that minimizes atmospheric background and Earth absorption effects in IceCube. We further exclude one source that lacks the hard X-ray flux estimate, as well as $19$ AGNs with fewer than three epochs in their MIR light curves (Section~\ref{sec:mir}). The final sample (hereafter the no-$\nu$ sample) consists of $387$ sources (i.e., excluding NGC 1068 and NGC 4151), of which $45$ are classified as beamed AGNs \citep{Koss2022}. As with NGC 1068 and NGC 4151 (see Section~\ref{sec:seyfert}), the intrinsic hard X-ray fluxes ($f_{14-150\ \mathrm{keV}}$) are taken from the $70$-month \textit{Swift}/BAT survey \citep{Ricci2017, Koss2022}. 

The intrinsic hard X-ray fluxes of the no-$\nu$ sample can be consistently compared with those of NGC 1068 or NGC 4151. Notably, several AGNs in the no-$\nu$ sample have fluxes comparable to or exceeding that of NGC 1068 (e.g., NGC 4388, NGC 5506), yet none are associated with IceCube neutrino events. Then, what makes NGC 1068 so special in terms of neutrino emission?

\subsection{MIR variational amplitude}
\label{sec:mir}
We calculate MIR variations ($\delta f$) using $W1$-band ($3.4\ \mathrm{\mu m}$) light curves from the \textit{Wide-field Infrared Survey Explorer} (\textit{WISE}). We retrieve both AllWISE \citep{WISE-DOI} and NEOWISE \citep{NEOWISE-DOI} observations\footnote{\url{https://irsa.ipac.caltech.edu/cgi-bin/Gator/nph-query}} via the python package \texttt{wise\_light\_curves}\footnote{\url{https://github.com/HC-Hwang/wise_light_curves?tab=readme-ov-file}} \citep{Hwang2020}. \textit{WISE} observes specific targets intensively every six months (hereafter, an epoch). The AllWISE epochs span MJD $55203$--$55593$, and NEOWISE epochs span MJD $56639$--$60523$. Within each epoch, multiple exposures are obtained over about a week. We adopt the profile-fit magnitudes (\texttt{w1mpro}), which are derived from point-spread-function (PSF) fitting of the source images. For bright sources with $W1\lesssim 8\ \mathrm{mag}$ (NGC 1068, NGC 4151, and NGC 5506), saturation can occur, leading to an overestimation of the measured brightness. To correct for this effect, we apply the saturation photometric bias corrections provided for profile-fit magnitudes\footnote{\url{https://irsa.ipac.caltech.edu/data/WISE/docs/release/NEOWISE/expsup/sec2_1civa.html}} to adjust the measurements for these three sources. 

To ensure data accuracy and quality, we take the following steps. First, we match source IDs between AllWISE and NEOWISE to confirm target identity. Next, we apply the selection criteria of \texttt{saa\_sep>0}, \texttt{moon\_masked=0000}, \texttt{qi\_fact>0.9}, and \texttt{cc\_flags=0000} for AllWISE observations.\footnote{\url{https://wise2.ipac.caltech.edu/docs/release/allwise/expsup/sec3_2.html}} For NEOWISE observations, an additional criterion \texttt{qual\_frame>0} is included.\footnote{\url{https://wise2.ipac.caltech.edu/docs/release/neowise/expsup/sec2_3.html}} Now we can obtain the MIR light curves from the reliable exposures of all epochs. For each epoch, if there are more than three reliable exposures, we calculate the median magnitude ($m_\mathrm{vega}$) of these exposures. The uncertainty of the median value is estimated via the normalized median absolute deviation of the reliable exposures. We calculate the MIR variation ($\delta f$) for each target if its MIR light curve has more than three epochs. To minimize the influence of the host galaxy, we calculate the MIR variation with $\delta f=(f_\mathrm{max}-f_\mathrm{min})/f_\mathrm{max}$, where $f=309.54\times10^{-m_\mathrm{vega}/2.5}\ \mathrm{[Jy]}$ is the MIR flux \citep{Wright2010}. The errors for $\delta f$ are obtained by error propagation. We note that since \textit{WISE} has a six-month cadence, the calculated $\delta f$ probes MIR variability on observed-frame timescales from $\sim180$ days to $\sim 14$ years. 

The MIR variations ($\delta f$) provide an effective probe of long-term fluctuations in the AGN central engines, on timescales matching the decade-long IceCube monitoring required to identify neutrino emitters. The dusty torus at sub-parsec scales absorbs X-ray/UV/optical photons and re-emits in the MIR, producing smoothed light curves \citep[e.g.,][]{Li2023} with time lags relative to X-ray/UV/optical light curves \citep[e.g.,][]{Lyu2019, Mandal2024}. For non-jetted AGNs in our sample, with bolometric luminosity of $10^{41.68}$--$10^{46.41}\ \mathrm{erg\ s^{-1}}$ \citep{Koss2022}, the expected rest-frame $W1$-optical time lags are $12$--$991$ days \citep[median $172$ days;][]{Mandal2024}, far shorter than the decade-long \textit{WISE} light curves. Thus, while short-term fluctuations of the AGN accretion disk and corona are largely absent in the reprocessed MIR emission, long-term variations are preserved in the MIR light curves. In jetted AGNs, the observed MIR emission may contain both torus and jet contributions. Therefore, MIR variations may at least partially reflect long-term jet fluctuations connected to neutrino production in blazars. 

\begin{figure}
\includegraphics[width=1.\linewidth]{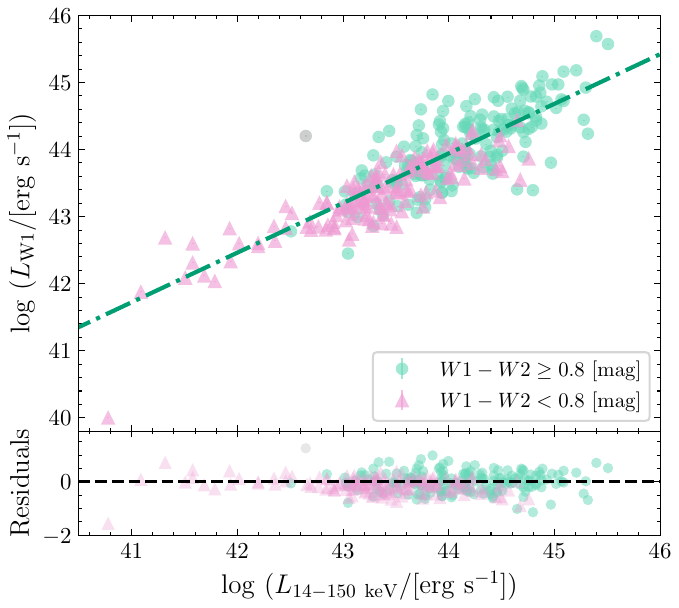}
\caption{The distributions of the $W1$ peak luminosity versus the 14–150 keV luminosity for Seyfert galaxies (including NGC 1068 and NGC 4151). Top panel: Green circles denote Seyferts with quasar-like MIR colors ($W1-W2\geq 0.8\ \mathrm{[mag]}$) in their brightest states, while purple triangles represent Seyferts with galaxy-like colors ($W1-W2< 0.8\ \mathrm{[mag]}$). Error bars show $1\sigma$ uncertainties (too small to be visible). The green dash-dotted line represents the best-fit relation for Seyferts with quasar-like MIR colors (excluding NGC 1068, indicated by the gray dot, which shows a significant MIR excess). Bottom panel: Residuals of the $W1$-band luminosity ($L_\mathrm{W1}$) relative to the expected values from the best-fit relation for Seyferts with quasar-like MIR colors($L_\mathrm{W1,exp}$; the green dash-dotted line). Both subsamples show small residuals, suggesting that the $W1$-band peak luminosity is a reliable tracer of the $14$–$150\ \mathrm{keV}$ luminosity. 
\label{fig1}}
\end{figure}

The potential host-galaxy contamination in $W1$ may dilute the AGN MIR variation. We assess the galaxy contribution using the MIR color diagnostics. The MIR color ($W1-W2$) with $W1-W2\geq 0.8\ \mathrm{[mag]}$ is adopted to select quasars whose MIR emission is largely free of host contamination \citep[e.g.,][]{Stern2012, Yan2013, Ichikawa2017}. We divide Seyferts into two subsamples: Seyferts with quasar-like MIR colors ($W1-W2\geq 0.8\ \mathrm{[mag]}$; $183$ sources) and Seyferts with galaxy-like MIR colors ($W1-W2< 0.8\ \mathrm{[mag]}$; $161$ sources). For Seyferts with quasar-like MIR colors, the $W1$-band luminosity is expected to correlate tightly with the $14$–$150\ \mathrm{keV}$ luminosity. We fit the relation between the $W1$-band peak luminosity and the hard X-ray luminosity for Seyferts with quasar-like MIR colors using the maximum likelihood method with the Markov Chain Monte Carlo sampling. We adopt the peak luminosity because, by definition, only the denominator ($f_{\mathrm{max}}$) in $\delta f$ is affected by the potential host galaxy emission. NGC 1068, which shows a significant MIR excess that may be caused by host galaxy contamination, is excluded in the fitting. The likelihood function is defined as $\ln \mathcal{L}=-0.5\sum[(L_\mathrm{W1,n}-L_\mathrm{model,W1,n})/\sigma_\mathrm{n}^2+\ln \sigma_\mathrm{n}^2]$, where $L_\mathrm{W1,n}$ is the observed $W1$-band peak luminosity and $L_\mathrm{model,W1,n}$ is the model prediction. The variance is $\sigma_\mathrm{n}^2=\sigma_\mathrm{n,err}^2+(\sigma_\mathrm{int}L_\mathrm{model,W1,n})^2$, with $\sigma_\mathrm{n,err}$ representing the measurement uncertainty of $L_\mathrm{W1,n}$ and $\sigma_\mathrm{int}L_\mathrm{model,W1,n}$ denoting the intrinsic scatter. Indeed, we find a tight relation (the green dash-dotted line in Figure~\ref{fig1}),
\begin{equation}\label{equ}
    \log \frac{L_{\mathrm{W1}}}{\mathrm{erg\ s^{-1}}}=0.74(\pm 0.05)\log \frac{L_{\mathrm{14-150\ keV}}}{10^{44}\ \mathrm{erg\ s^{-1}}} +43.94(\pm 0.03)
\end{equation}. 
For Seyferts with galaxy-like MIR colors, it is unclear whether the host galaxy significantly contributes to the $W1$-band peak luminosity. To address this, we calculate the residuals between the best-fit relation for Seyferts with quasar-like MIR colors ($L_\mathrm{W1,exp}$; Eq.~\ref{equ}) and the observed $W1$-band luminosities ($L_\mathrm{W1}$), i.e., $\mathrm{Residuals}=\mathrm{log}({L_\mathrm{W1}}/{L_\mathrm{W1,exp}})$ (the bottom panel of Figure~\ref{fig1}). The median residual is $0.01$ for Seyferts with quasar-like MIR colors and $-0.09$ for those with galaxy-like MIR colors. This small difference indicates that the $W1$-band peak emission serves as a reliable tracer of the $14$–$150\ \mathrm{keV}$ emission for both subsamples. Hence, we conclude that, for most Seyferts, host-galaxy contamination should not significantly bias our MIR variability ($\delta f$) calculations. 

\section{Results} \label{sec:res}

\begin{figure}
\includegraphics[width=1.\linewidth]{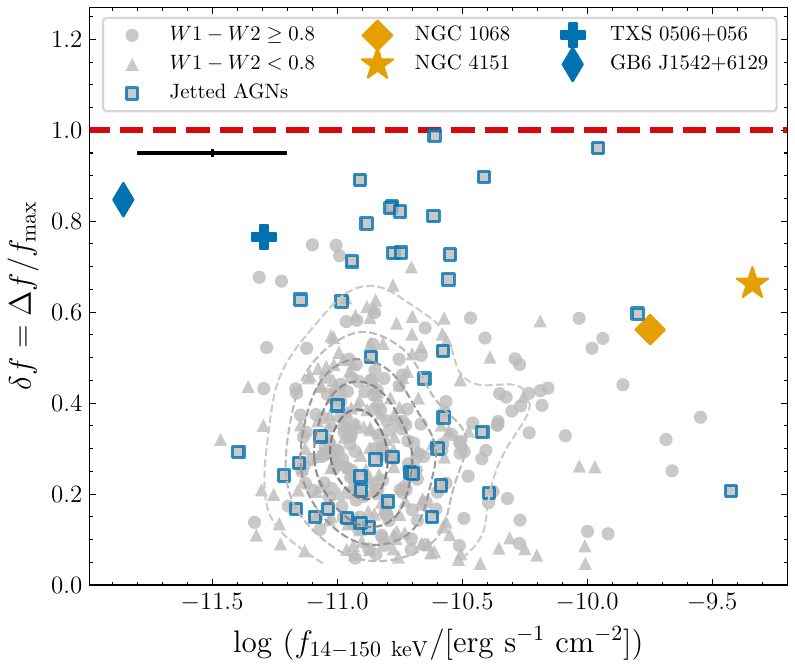}
\caption{Hard X-ray fluxes ($f_{14-150\ \mathrm{keV}}$) and MIR variations ($\delta f$) for the neutrino emitters and the no-$\nu$ sample. Gray circles and triangles represent Seyferts with quasar-like MIR colors ($W1-W2\geq 0.8\ \mathrm{[mag]}$) and Seyferts with galaxy-like MIR colors ($W1-W2< 0.8\ \mathrm{[mag]}$) in the no-$\nu$ sample, respectively. Gray squares with blue borders denote jetted AGNs in the no-$\nu$ sample. Gray dashed curves show the number density of the no-$\nu$ sample. Yellow and blue filled symbols denote Seyfert and blazar neutrino emitters, respectively. The solid black bars in the upper-left corner show systematic uncertainties for $14$--$150\ \mathrm{keV}$ flux and MIR variation. Seyfert neutrino emitters exhibit either high hard X-ray fluxes or significant MIR variations.
\label{fig2}}
\end{figure}

Figure~\ref{fig2} shows the MIR variations ($\delta f$) and hard X-ray fluxes ($f_{14-150\ \mathrm{keV}}$) for the neutrino emitters and the no-$\nu$ sample. Several X-ray bright  Seyferts (NGC 4388, NGC 5506, and Z164-19) exhibit intrinsic hard X-ray fluxes comparable to that of NGC 1068 but are not associated with neutrino detections. Compared with these Seyferts, NGC 1068 is unique in terms of its relatively large MIR variations.\footnote{The intrinsic MIR variation of NGC 1068 would be even larger than we obtained if its host galaxy contamination is properly subtracted.} Interestingly, the other candidate neutrino emitter, NGC 4151, also shows both a large hard X-ray flux and a large MIR variation. Hence, we speculate that the large hard X-ray flux and violent long-term fluctuations of the central engine are the two critical factors for neutrino production in Seyfert galaxies. We caution, however, that this conclusion is based on a limited number of neutrino detections and requires confirmation by future IceCube observations. 

Our results may guide the analysis of the association between neutrinos and EM counterparts. To this end, we list in Table~\ref{tab:candi} the Seyferts (i.e., without blazars) whose hard X-ray fluxes and MIR variations are comparable to those of NGC 1068. The criteria are $\log (f_{14-150\ \mathrm{keV}}/[\mathrm{erg\ s^{-1}\ cm^{-2}}])>-10.5$ and $\delta f>0.4$. If our speculation is correct, stacking analyses of existing IcuCube data or future observations may reveal neutrino signals from these sources. While preparing to submit the manuscript, we note that two neutrino alerts have been reported in spatial coincidence with NGC 7469 \citep{Sommani2025, Zegarelli2025}. The hard X-ray flux of NGC 7469 is modest ($5.58\times 10^{-11}\ \mathrm{erg\ s^{-1}\ cm^{-2}}$), which is fainter than a significant fraction of BASS DR2 Seyferts. Nevertheless, its MIR variation ($\delta f$ is $0.42$) is larger than that of most Seyferts with comparable or higher fluxes. The possible association of NGC 7469 with neutrino events further underscores the potential importance of long-term AGN central engine fluctuations in producing neutrinos. 

If confirmed, our speculation that long-term AGN central engine variability is a key factor in neutrino production would provide important new insight into the origin of neutrinos in Seyferts. For instance, scenarios invoking host galaxy origins \citep[e.g.,][]{Sridhar2024}, which are not directly linked to central engine fluctuations, are unlikely to dominate the IceCube neutrinos observed in Seyfert galaxies. In addition, blazar neutrino emitters also exhibit significant MIR variations, suggesting a potential correlation with long-term jet fluctuations.

\section{Discussions}\label{sec:discus}
Here, we discuss the implications of our results (if confirmed by future neutrino observations) for current models of neutrino production in AGNs. The AGN accretion flow and corona have long been considered promising sites for particle acceleration and neutrino production \citep[e.g.,][]{Begelman1990, Stecker1991, Kimura2019-AGN-IRAF, Murase2020-AGN-corona}. The high optical thickness of the corona to GeV--TeV gamma-rays naturally explains the absence of strong gamma-ray counterparts in Seyfert neutrino emitters. The mechanisms of particle acceleration in the corona are diverse. On one hand, \citet{Inoue2019-AGN-corona} propose that diffuse shock acceleration in a homogeneous corona is the most effective process for producing the observed neutrinos in NGC 1068. This view is supported by radio observations of two nearby Seyfert galaxies \citep{Inoue2019-AGN-corona}; the radio emission is likely co-spatial with the X-ray according to ALMA and \textit{Swift} observations \citep{Ricci2023}. The detected radio synchrotron emission by ALMA indicates a coronal magnetic field strength of $\sim 10\ \mathrm{G}$ \citep{Inoue2018}, which is too weak to support efficient stochastic or reconnection acceleration \citep{Inoue2019-AGN-corona}. On the other hand, \citet{Gutierrez2021-AGN-IRAF} suggest that the radio emission can arise from the outer regions of an inhomogeneous corona with decreasing magnetic fields from inner to outer regions. Thus, in the inner regions with strong magnetic fields, stochastic acceleration in magnetized turbulence \citep[e.g.,][]{Murase2020-AGN-corona, Eichmann2022-AGN_corona, Fiorillo2024a-AGN-corona, Lemoine2024-AGN-corona} and magnetic reconnection \citep[e.g.,][]{Kheirandish2021-AGN-corona, Fiorillo2024b-AGN-corona} may still operate effectively. Regardless of the specific acceleration mechanism, coronal scenarios generally predict a connection between X-ray and neutrino fluxes \citep[e.g.,][]{Kheirandish2021-AGN-corona, Kun2024-AGN-blazar}. Our results emphasize that the corona should also exhibit strong fluctuations on long-term timescales. 

The central engines of Seyfert galaxies can fluctuate due to several mechanisms. First-principle particle-in-cell simulations \citep{Fiorillo2024a-AGN-corona, Mbarek2024-AGN-corona, Fiorillo2025-AGN-corona} show that strong magnetic turbulence is required to accelerate particles and produce the observed IceCube neutrino luminosity. Magnetic turbulence in the AGN accretion disk is often invoked as an essential channel for dissipating accretion power into thermal energy \citep[e.g.,][]{Balbus2003} and may also power hot coronae \citep[e.g.,][]{Merloni2001}. Moreover, numerical magnetohydrodynamic simulations reveal that the magnetic turbulence in the accretion disk can drive thermal fluctuations, possibly leading to the observed temporal variability \citep{Jiang2013}. Therefore, the strong magnetic turbulence required for the stochastic mechanism explaining the observed neutrino fluxes in Seyfert galaxies may lead to significant long-term thermal fluctuations in the accretion disk, resulting in considerable UV/optical variability and manifesting as MIR variations. In addition, \citet{Fiorillo2024b-AGN-corona} propose that magnetic reconnection in the AGN corona could also be the dominant mechanism for the observed neutrinos. The magnetic reconnection is likely to induce multiple short-time flares \citep[hours to days; e.g.,][]{Matteo1998, Fiorillo2024b-AGN-corona}, but to account for the MIR variations in neutrino emitters, the reconnection rates may fluctuate strongly over long timescales. 

Coronal emission may also be modeled as an advection-dominated hot accretion flow, i.e., RIAFs \citep[for a review, see][]{Yuan2014}. The RIAFs have been proposed as a potential site for particle acceleration \citep[e.g.,][]{Begelman1990, Stecker1991, Kimura2019-AGN-IRAF, Kimura2021-AGN-IRAF, Gutierrez2021-AGN-IRAF, Inoue2024}. Regardless of whether the RIAFs can explain the observed neutrinos in NGC 1068 and NGC 4151, the advection-dominated accretion disk can induce large amplitude variability \citep[e.g.,][]{Manmoto1996}. 

Fast outflows have been observed in AGNs \citep[e.g.,][]{Veilleux2005, King2015, Laha2021}. Particles can be accelerated by the outflows near central supermassive black holes \citep{Inoue2022-AGN-interact}, as well as by collisions between these outflows and surrounding materials \citep[e.g.,][]{Mou2021, Inoue2022-AGN-interact, Huang2024-AGN-interact}. Based on our results showing large MIR variations on timescales of years for neutrino emitters, it appears that the interactions between outflows and interstellar medium on large scales (dozens of parsecs) around the nuclear core \citep[e.g.,][]{Fang2023-AGN-interact, Peretti2023-AGN-interact, Ehlert2025-AGN-interact} may not be the main mechanism for the observed neutrinos in Seyfert galaxies. Instead, the possible interaction between outflows and gaseous clouds should occur on sub-pc scales. Moreover, given the estimated destruction lifetime of the clouds is relatively short \citep[$\sim 30$--$60\ \mathrm{s}$;] []{Huang2024-AGN-interact}, it is reasonable to require that the outflow be episodic, ensuring enough time for the clouds to reform. The episodic outflow likely arises from the fluctuations of the central engine, which would be accompanied by multi-wavelength variability. 

In addition to the interactions of winds/outflows with the surrounding material, \citet{Muller2020-AGN-interact} propose that the interactions between broad line region clouds and the AGN accretion disk can produce neutrinos. If this process is universal in AGNs, it is unclear why neutrinos are strong in NGC 1068 rather than in other Seyferts. 

The evolution and explosions of stars and compact objects embedded in the AGN accretion disk have attracted considerable discussion \citep[e.g.,][]{Cheng1999, Cantiello2021}. The special environment of the AGN accretion disk serves as a breeding ground for multi-messenger radiation. According to \citet{Tagawa2023-AGN-AMS}, jetted stellar-mass black holes in the accretion disk can produce intense EM emission, as well as the observed neutrinos in NGC 1068. Furthermore, phenomena such as stellar explosions \citep{Zhu2021a-AGN-AMS}, choked gamma-ray bursts \citep{Zhu2021b-AGN-AMS}, and merging stellar-mass black holes \citep[e.g.,][]{Zhu2024-AGN-AMS, Ma2024-AGN-AMS} in the AGN accretion disk are also expected to produce neutrinos. However, the relatively low level of UV/optical radiation from the jet processes \citep[$\lesssim10^{40}\ \mathrm{erg\ s^{-1}}$;][]{Tagawa2023-AGN-AMS} or explosions of stellar objects in AGN accretion disks are unlikely to dominate over the emission from the AGN accretion disk itself and induce MIR variability. While there may be a sufficient number of jetted stellar-mass black holes \citep[e.g.,][]{Zhou2024} or explosive objects that could produce emission comparable to that from the AGN accretion disk, the large number of these objects is unlikely to result in significant variations because these objects can hardly vary coherently. Therefore, the significant MIR variations in the neutrino emitters suggest that the jet processes or explosions of stellar objects in the AGN accretion disk may not be the primary mechanisms responsible for neutrino emission. 

In blazars, the observed MIR emission may originate from both the dusty torus and the relativistic jet. Let us first consider the possibility that the observed MIR emission in blazar neutrino emitters is dominated by dusty tori. Then, like Seyferts, the intrinsic MIR variations would indicate long-term central engine fluctuations. Several studies have suggested that blazars and Seyferts may share a common neutrino origin \citep[i.e., the corona][]{Kun2024-AGN-blazar, Zathul2024-AGN-blazar, Yang2025-AGN-blazar}. However, \citet{Fiorillo2025-AGN-blazar} argue that coronal neutrino production in TXS 0506+056 is insufficient to explain the observations. Instead, neutrino emission in TXS 0506+056 is likely related to the relativistic jet. Under the assumption of the torus origin, the large MIR variation in TXS 0506+056 suggests that fluctuations in the central engine may propagate into the jet, effectively accelerating cosmic rays. Alternatively, the observed MIR emission in blazars is dominated by jets and directly probes jet fluctuations. 

\begin{figure}
\includegraphics[width=1.\linewidth]{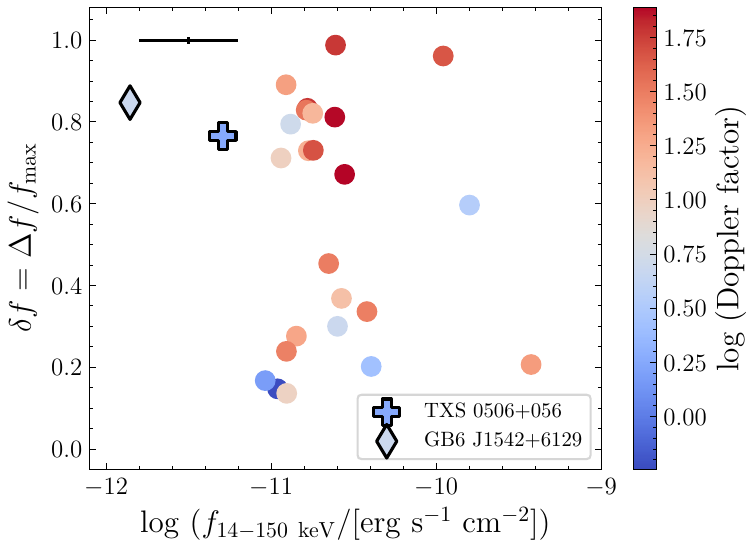}
\caption{The Doppler factors for blazars measured by \citet{Homan2021}. Note that only $24$ out of $45$ blazars in the control sample have Doppler factor measurements. The color indicates the Doppler factor values. Blazars with higher $\delta f$ tend to have larger Doppler factors, suggesting that the significant apparent MIR variations in blazars without neutrino detection may be attributed to the relativistic beaming effect. 
\label{fig3}}
\end{figure}

The jet origin of the observed MIR variations in blazars depends on both the Doppler effect and the intrinsic jet MIR variations. The Doppler factor indicates the amplification of intrinsic variability due to the beaming effect. Figure~\ref{fig3} shows the Doppler factors \citep[estimated by][]{Homan2021} for blazars in the no-$\nu$ sample and for blazar neutrino emitters. Blazars in the no-$\nu$ sample that have higher fluxes and apparent MIR variations than TXS 0506+056 and GB6 J1542+6129 tend to have large Doppler factors. This conclusion remains unchanged if one adopts Doppler factors measured by another method of \cite{Liodakis2018}. Thus, their intrinsic MIR variations are smaller than the observed ones. Then, we speculate that the blazar neutrino emitters also exhibit large \textit{intrinsic} long-term jet fluctuations.

\section{Conclusions} \label{sec:conclusion}
In this work, we have compared the intrinsic hard X-ray fluxes and MIR variations of high-confidence neutrino emitters with those of the no-$\nu$ sample from BASS DR2. We have found that neutrino emitters, including both Seyfert galaxies and blazars, exhibit significant \textit{intrinsic} MIR variations. We speculate that, in addition to intrinsic hard X-ray flux, long-term fluctuations of the central engine may also play a crucial role in producing neutrinos. This hypothesis can guide future neutrino-EM counterpart association studies. Following our speculation, we have provided a list of Seyferts that may be neutrino emitters, which can be verified through stacking analyses or future observations. If confirmed, our speculation has several implications for the neutrino production mechanisms in AGNs. First, the origin of neutrinos is unlikely to be dominated by host galaxy processes or large-scale (dozens of parsecs) interactions between the winds/outflows and the surrounding interstellar medium. Second, in the coronal scenario, energy dissipation through magnetic turbulence and reconnection should be efficient and highly time-dependent on long timescales. Third, the jets may be intrinsically unstable over long timescales for blazar neutrino emitters. Fourth, in models involving interactions between AGN winds/outflows and surrounding materials, these winds/outflows are likely episodic rather than continuous. Finally, the jet processes or explosions of stellar objects in the AGN accretion disk are unlikely to be the main source of the observed neutrinos. 

\begin{deluxetable}{ccc}
\label{tab:candi}
\tablewidth{0pt}
\tablecaption{Seyferts with large hard X-ray fluxes and MIR variability}
\tablehead{
\colhead{Name} & \colhead{$f_{14-150\ \mathrm{keV}}$} & \colhead{$(f_{\mathrm{max}}-f_{\mathrm{min}})/f_{\mathrm{max}}$}\\
\colhead{} & \colhead{$\mathrm{log\ [erg\ s^{-1}\ cm^{-2}]}$} & \colhead{}}
\startdata
NGC 4151$^{a}$ & $-9.34$ & $0.66$ \\
NGC 1068$^{b}$ & $-9.74$ & $0.56$ \\
\hline
Mrk 348 & $-9.86$ & $0.44$\\
UGC 3374 & $-9.94$ & $0.54$\\
3C 111 & $-9.98$ & $0.52$\\
NGC 5252 & $-10.03$ & $0.59$\\
NGC 1275 & $-10.16$ & $0.43$\\
3C 382 & $-10.19$ & $0.43$\\
2MASX J20145928+2523010 & $-10.19$ & $0.58$\\
NGC 4992 & $-10.23$ & $0.42$\\
NGC 7469$^{c}$ & $-10.25$ & $0.42$\\
Mrk 1040 & $-10.27$ & $0.49$\\
LEDA 168563 & $-10.29$ & $0.50$\\
Fairall 272 & $-10.32$ & $0.41$\\
LEDA 138501 & $-10.35$ & $0.42$\\
LEDA 86269 & $-10.39$ & $0.50$\\
Mrk 79 & $-10.41$ & $0.54$\\
Z121-75 & $-10.47$ & $0.59$\\
\enddata
\tablecomments{a. neutrino association with a significance level of $2.9\sigma$ is reported by \cite{Abbasi2025-1068-4151}; b. neutrino association with a significance level of $4.2\sigma$ NGC 1068 is reported by \cite{IceCube2022-1068}; c. a possible neutrino association is claimed by \cite{Sommani2025} and \cite{Zegarelli2025}.}
\end{deluxetable}

\begin{acknowledgments}
We thank the anonymous referee for his/her constructive comments that improved the manuscript. We acknowledge Manqi Fu for her helpful suggestions on visualization and Junfeng Wang for his helpful discussions. S.Y.Z. and M.Y.S. acknowledge support from the National Natural Science Foundation of China (NSFC-12322303). G.B.M. acknowledges support from the National Natural Science Foundation of China (NSFC-12473013). D.B.L. acknowledges support from the National Natural Science Foundation of China (NSFC-12273005). T.L. acknowledges support from the National Natural Science Foundation of China (under grants NSFC-12173031, NSFC-12494572, and NSFC-12221003) and the Fund of National Key Laboratory of Plasma Physics (No. 6142A04240201). Y.Q.X. acknowledges support from the National Natural Science Foundation of China (NSFC-12025303 and NSFC-12393814). This research was also supported by the science research grants from the China Manned Space Project with No.~CMS-CSST-2025-A13, and the Fundamental Research Funds for the Central Universities (20720240152). This publication makes use of data products from the Wide-field Infrared Survey Explorer, which is a joint project of the University of California, Los Angeles, and the Jet Propulsion Laboratory/California Institute of Technology, funded by the National Aeronautics and Space Administration. This publication also makes use of data products from NEOWISE, which is a project of the Jet Propulsion Laboratory/California Institute of Technology, funded by the Planetary Science Division of the National Aeronautics and Space Administration. 
\end{acknowledgments}

\facilities{WISE}
\software{astropy \citep{Astropy2013, Astropy2018, Astropy2022}, Matplotlib \citep{matplotlib}, Numpy \citep{2020NumPy-Array}, Scipy \citep{2020SciPy-NMeth}, wise\_light\_curves \citep{Hwang2020}}

\newpage
\bibliography{reference}{}

@ARTICLE{Astropy2022,
       author = {{Astropy Collaboration} and {Price-Whelan}, Adrian M. and {Lim}, Pey Lian and {Earl}, Nicholas and {Starkman}, Nathaniel and {Bradley}, Larry and {Shupe}, David L. and {Patil}, Aarya A. and {Corrales}, Lia and {Brasseur}, C.~E. and {N{\"o}the}, Maximilian and {Donath}, Axel and {Tollerud}, Erik and {Morris}, Brett M. and {Ginsburg}, Adam and {Vaher}, Eero and {Weaver}, Benjamin A. and {Tocknell}, James and {Jamieson}, William and {van Kerkwijk}, Marten H. and {Robitaille}, Thomas P. and {Merry}, Bruce and {Bachetti}, Matteo and {G{\"u}nther}, H. Moritz and {Aldcroft}, Thomas L. and {Alvarado-Montes}, Jaime A. and {Archibald}, Anne M. and {B{\'o}di}, Attila and {Bapat}, Shreyas and {Barentsen}, Geert and {Baz{\'a}n}, Juanjo and {Biswas}, Manish and {Boquien}, M{\'e}d{\'e}ric and {Burke}, D.~J. and {Cara}, Daria and {Cara}, Mihai and {Conroy}, Kyle E. and {Conseil}, Simon and {Craig}, Matthew W. and {Cross}, Robert M. and {Cruz}, Kelle L. and {D'Eugenio}, Francesco and {Dencheva}, Nadia and {Devillepoix}, Hadrien A.~R. and {Dietrich}, J{\"o}rg P. and {Eigenbrot}, Arthur Davis and {Erben}, Thomas and {Ferreira}, Leonardo and {Foreman-Mackey}, Daniel and {Fox}, Ryan and {Freij}, Nabil and {Garg}, Suyog and {Geda}, Robel and {Glattly}, Lauren and {Gondhalekar}, Yash and {Gordon}, Karl D. and {Grant}, David and {Greenfield}, Perry and {Groener}, Austen M. and {Guest}, Steve and {Gurovich}, Sebastian and {Handberg}, Rasmus and {Hart}, Akeem and {Hatfield-Dodds}, Zac and {Homeier}, Derek and {Hosseinzadeh}, Griffin and {Jenness}, Tim and {Jones}, Craig K. and {Joseph}, Prajwel and {Kalmbach}, J. Bryce and {Karamehmetoglu}, Emir and {Ka{\l}uszy{\'n}ski}, Miko{\l}aj and {Kelley}, Michael S.~P. and {Kern}, Nicholas and {Kerzendorf}, Wolfgang E. and {Koch}, Eric W. and {Kulumani}, Shankar and {Lee}, Antony and {Ly}, Chun and {Ma}, Zhiyuan and {MacBride}, Conor and {Maljaars}, Jakob M. and {Muna}, Demitri and {Murphy}, N.~A. and {Norman}, Henrik and {O'Steen}, Richard and {Oman}, Kyle A. and {Pacifici}, Camilla and {Pascual}, Sergio and {Pascual-Granado}, J. and {Patil}, Rohit R. and {Perren}, Gabriel I. and {Pickering}, Timothy E. and {Rastogi}, Tanuj and {Roulston}, Benjamin R. and {Ryan}, Daniel F. and {Rykoff}, Eli S. and {Sabater}, Jose and {Sakurikar}, Parikshit and {Salgado}, Jes{\'u}s and {Sanghi}, Aniket and {Saunders}, Nicholas and {Savchenko}, Volodymyr and {Schwardt}, Ludwig and {Seifert-Eckert}, Michael and {Shih}, Albert Y. and {Jain}, Anany Shrey and {Shukla}, Gyanendra and {Sick}, Jonathan and {Simpson}, Chris and {Singanamalla}, Sudheesh and {Singer}, Leo P. and {Singhal}, Jaladh and {Sinha}, Manodeep and {Sip{\H{o}}cz}, Brigitta M. and {Spitler}, Lee R. and {Stansby}, David and {Streicher}, Ole and {{\v{S}}umak}, Jani and {Swinbank}, John D. and {Taranu}, Dan S. and {Tewary}, Nikita and {Tremblay}, Grant R. and {de Val-Borro}, Miguel and {Van Kooten}, Samuel J. and {Vasovi{\'c}}, Zlatan and {Verma}, Shresth and {de Miranda Cardoso}, Jos{\'e} Vin{\'\i}cius and {Williams}, Peter K.~G. and {Wilson}, Tom J. and {Winkel}, Benjamin and {Wood-Vasey}, W.~M. and {Xue}, Rui and {Yoachim}, Peter and {Zhang}, Chen and {Zonca}, Andrea and {Astropy Project Contributors}},
        title = "{The Astropy Project: Sustaining and Growing a Community-oriented Open-source Project and the Latest Major Release (v5.0) of the Core Package}",
      journal = {\apj},
         year = 2022,
        month = aug,
       volume = {935},
       number = {2},
          eid = {167},
        pages = {167},
          doi = {10.3847/1538-4357/ac7c74},
archivePrefix = {arXiv},
       eprint = {2206.14220},
 primaryClass = {astro-ph.IM},
       adsurl = {https://ui.adsabs.harvard.edu/abs/2022ApJ...935..167A}
}

@ARTICLE{Astropy2018,
       author = {{Astropy Collaboration} and {Price-Whelan}, A.~M. and {Sip{\H{o}}cz}, B.~M. and {G{\"u}nther}, H.~M. and {Lim}, P.~L. and {Crawford}, S.~M. and {Conseil}, S. and {Shupe}, D.~L. and {Craig}, M.~W. and {Dencheva}, N. and {Ginsburg}, A. and {VanderPlas}, J.~T. and {Bradley}, L.~D. and {P{\'e}rez-Su{\'a}rez}, D. and {de Val-Borro}, M. and {Aldcroft}, T.~L. and {Cruz}, K.~L. and {Robitaille}, T.~P. and {Tollerud}, E.~J. and {Ardelean}, C. and {Babej}, T. and {Bach}, Y.~P. and {Bachetti}, M. and {Bakanov}, A.~V. and {Bamford}, S.~P. and {Barentsen}, G. and {Barmby}, P. and {Baumbach}, A. and {Berry}, K.~L. and {Biscani}, F. and {Boquien}, M. and {Bostroem}, K.~A. and {Bouma}, L.~G. and {Brammer}, G.~B. and {Bray}, E.~M. and {Breytenbach}, H. and {Buddelmeijer}, H. and {Burke}, D.~J. and {Calderone}, G. and {Cano Rodr{\'\i}guez}, J.~L. and {Cara}, M. and {Cardoso}, J.~V.~M. and {Cheedella}, S. and {Copin}, Y. and {Corrales}, L. and {Crichton}, D. and {D'Avella}, D. and {Deil}, C. and {Depagne}, {\'E}. and {Dietrich}, J.~P. and {Donath}, A. and {Droettboom}, M. and {Earl}, N. and {Erben}, T. and {Fabbro}, S. and {Ferreira}, L.~A. and {Finethy}, T. and {Fox}, R.~T. and {Garrison}, L.~H. and {Gibbons}, S.~L.~J. and {Goldstein}, D.~A. and {Gommers}, R. and {Greco}, J.~P. and {Greenfield}, P. and {Groener}, A.~M. and {Grollier}, F. and {Hagen}, A. and {Hirst}, P. and {Homeier}, D. and {Horton}, A.~J. and {Hosseinzadeh}, G. and {Hu}, L. and {Hunkeler}, J.~S. and {Ivezi{\'c}}, {\v{Z}}. and {Jain}, A. and {Jenness}, T. and {Kanarek}, G. and {Kendrew}, S. and {Kern}, N.~S. and {Kerzendorf}, W.~E. and {Khvalko}, A. and {King}, J. and {Kirkby}, D. and {Kulkarni}, A.~M. and {Kumar}, A. and {Lee}, A. and {Lenz}, D. and {Littlefair}, S.~P. and {Ma}, Z. and {Macleod}, D.~M. and {Mastropietro}, M. and {McCully}, C. and {Montagnac}, S. and {Morris}, B.~M. and {Mueller}, M. and {Mumford}, S.~J. and {Muna}, D. and {Murphy}, N.~A. and {Nelson}, S. and {Nguyen}, G.~H. and {Ninan}, J.~P. and {N{\"o}the}, M. and {Ogaz}, S. and {Oh}, S. and {Parejko}, J.~K. and {Parley}, N. and {Pascual}, S. and {Patil}, R. and {Patil}, A.~A. and {Plunkett}, A.~L. and {Prochaska}, J.~X. and {Rastogi}, T. and {Reddy Janga}, V. and {Sabater}, J. and {Sakurikar}, P. and {Seifert}, M. and {Sherbert}, L.~E. and {Sherwood-Taylor}, H. and {Shih}, A.~Y. and {Sick}, J. and {Silbiger}, M.~T. and {Singanamalla}, S. and {Singer}, L.~P. and {Sladen}, P.~H. and {Sooley}, K.~A. and {Sornarajah}, S. and {Streicher}, O. and {Teuben}, P. and {Thomas}, S.~W. and {Tremblay}, G.~R. and {Turner}, J.~E.~H. and {Terr{\'o}n}, V. and {van Kerkwijk}, M.~H. and {de la Vega}, A. and {Watkins}, L.~L. and {Weaver}, B.~A. and {Whitmore}, J.~B. and {Woillez}, J. and {Zabalza}, V. and {Astropy Contributors}},
        title = "{The Astropy Project: Building an Open-science Project and Status of the v2.0 Core Package}",
      journal = {\aj},
         year = 2018,
        month = sep,
       volume = {156},
       number = {3},
          eid = {123},
        pages = {123},
          doi = {10.3847/1538-3881/aabc4f},
archivePrefix = {arXiv},
       eprint = {1801.02634},
 primaryClass = {astro-ph.IM},
       adsurl = {https://ui.adsabs.harvard.edu/abs/2018AJ....156..123A}
}

@ARTICLE{Astropy2013,
       author = {{Astropy Collaboration} and {Robitaille}, Thomas P. and
         {Tollerud}, Erik J. and {Greenfield}, Perry and {Droettboom}, Michael and
         {Bray}, Erik and {Aldcroft}, Tom and {Davis}, Matt and
         {Ginsburg}, Adam and {Price-Whelan}, Adrian M. and
         {Kerzendorf}, Wolfgang E. and {Conley}, Alexander and {Crighton}, Neil and
         {Barbary}, Kyle and {Muna}, Demitri and {Ferguson}, Henry and
         {Grollier}, Fr{\'e}d{\'e}ric and {Parikh}, Madhura M. and
         {Nair}, Prasanth H. and {Unther}, Hans M. and {Deil}, Christoph and
         {Woillez}, Julien and {Conseil}, Simon and {Kramer}, Roban and
         {Turner}, James E.~H. and {Singer}, Leo and {Fox}, Ryan and
         {Weaver}, Benjamin A. and {Zabalza}, Victor and {Edwards}, Zachary I. and
         {Azalee Bostroem}, K. and {Burke}, D.~J. and {Casey}, Andrew R. and
         {Crawford}, Steven M. and {Dencheva}, Nadia and {Ely}, Justin and
         {Jenness}, Tim and {Labrie}, Kathleen and {Lim}, Pey Lian and
         {Pierfederici}, Francesco and {Pontzen}, Andrew and {Ptak}, Andy and
         {Refsdal}, Brian and {Servillat}, Mathieu and {Streicher}, Ole},
        title = "{Astropy: A community Python package for astronomy}",
      journal = {\aap},
         year = "2013",
        month = "Oct",
       volume = {558},
          eid = {A33},
        pages = {A33},
          doi = {10.1051/0004-6361/201322068},
archivePrefix = {arXiv},
       eprint = {1307.6212},
 primaryClass = {astro-ph.IM},
       adsurl = {https://ui.adsabs.harvard.edu/abs/2013A&A...558A..33A}
}

@ARTICLE{2020NumPy-Array,
  author  = {Harris, Charles R. and Millman, K. Jarrod and van der Walt, Stéfan J and Gommers, Ralf and Virtanen, Pauli and Cournapeau, David and Wieser, Eric and Taylor, Julian and Berg, Sebastian and Smith, Nathaniel J. and Kern, Robert and Picus, Matti and Hoyer, Stephan and van Kerkwijk, Marten H. and Brett, Matthew and Haldane, Allan and Fernández del Río, Jaime and Wiebe, Mark and Peterson, Pearu and Gérard-Marchant, Pierre and Sheppard, Kevin and Reddy, Tyler and Weckesser, Warren and Abbasi, Hameer and Gohlke, Christoph and Oliphant, Travis E.},
  title   = {Array programming with {NumPy}},
  journal = {Nature},
  year    = {2020},
  volume  = {585},
  pages   = {357–362},
  doi     = {10.1038/s41586-020-2649-2}
}

@ARTICLE{2020SciPy-NMeth,
  author  = {Virtanen, Pauli and Gommers, Ralf and Oliphant, Travis E. and
            Haberland, Matt and Reddy, Tyler and Cournapeau, David and
            Burovski, Evgeni and Peterson, Pearu and Weckesser, Warren and
            Bright, Jonathan and {van der Walt}, St{\'e}fan J. and
            Brett, Matthew and Wilson, Joshua and Millman, K. Jarrod and
            Mayorov, Nikolay and Nelson, Andrew R. J. and Jones, Eric and
            Kern, Robert and Larson, Eric and Carey, C J and
            Polat, {\.I}lhan and Feng, Yu and Moore, Eric W. and
            {VanderPlas}, Jake and Laxalde, Denis and Perktold, Josef and
            Cimrman, Robert and Henriksen, Ian and Quintero, E. A. and
            Harris, Charles R. and Archibald, Anne M. and
            Ribeiro, Ant{\^o}nio H. and Pedregosa, Fabian and
            {van Mulbregt}, Paul and {SciPy 1.0 Contributors}},
  title   = {{{SciPy} 1.0: Fundamental Algorithms for Scientific
            Computing in Python}},
  journal = {Nature Methods},
  year    = {2020},
  volume  = {17},
  pages   = {261--272},
  adsurl  = {https://rdcu.be/b08Wh},
  doi     = {10.1038/s41592-019-0686-2},
}

@ARTICLE{matplotlib,
       author = {{Hunter}, John D.},
        title = "{Matplotlib: A 2D Graphics Environment}",
      journal = {Computing in Science and Engineering},
         year = 2007,
        month = may,
       volume = {9},
       number = {3},
        pages = {90-95},
          doi = {10.1109/MCSE.2007.55},
       adsurl = {https://ui.adsabs.harvard.edu/abs/2007CSE.....9...90H}
}

@ARTICLE{Koss2022,
       author = {{Koss}, Michael J. and {Ricci}, Claudio and {Trakhtenbrot}, Benny and {Oh}, Kyuseok and {den Brok}, Jakob S. and {Mej{\'\i}a-Restrepo}, Julian E. and {Stern}, Daniel and {Privon}, George C. and {Treister}, Ezequiel and {Powell}, Meredith C. and {Mushotzky}, Richard and {Bauer}, Franz E. and {Ananna}, Tonima T. and {Balokovi{\'c}}, Mislav and {B{\"a}r}, Rudolf E. and {Becker}, George and {Bessiere}, Patricia and {Burtscher}, Leonard and {Caglar}, Turgay and {Congiu}, Enrico and {Evans}, Phil and {Harrison}, Fiona and {Heida}, Marianne and {Ichikawa}, Kohei and {Kamraj}, Nikita and {Lamperti}, Isabella and {Pacucci}, Fabio and {Ricci}, Federica and {Riffel}, Rog{\'e}rio and {Rojas}, Alejandra F. and {Schawinski}, Kevin and {Temple}, Matthew J. and {Urry}, C. Megan and {Veilleux}, Sylvain and {Williams}, Jonathan},
        title = "{BASS. XXII. The BASS DR2 AGN Catalog and Data}",
      journal = {\apjs},
         year = 2022,
        month = jul,
       volume = {261},
       number = {1},
          eid = {2},
        pages = {2},
          doi = {10.3847/1538-4365/ac6c05},
archivePrefix = {arXiv},
       eprint = {2207.12432},
 primaryClass = {astro-ph.GA},
       adsurl = {https://ui.adsabs.harvard.edu/abs/2022ApJS..261....2K}
}

@ARTICLE{IceCube2013,
       author = {{IceCube Collaboration}},
        title = "{Evidence for High-Energy Extraterrestrial Neutrinos at the IceCube Detector}",
      journal = {Science},
         year = 2013,
        month = nov,
       volume = {342},
       number = {6161},
          eid = {1242856},
        pages = {1242856},
          doi = {10.1126/science.1242856},
archivePrefix = {arXiv},
       eprint = {1311.5238},
 primaryClass = {astro-ph.HE},
       adsurl = {https://ui.adsabs.harvard.edu/abs/2013Sci...342E...1I}
}

@ARTICLE{IceCube2018a-TXS,
       author = {{IceCube Collaboration} and {Aartsen}, M.~G. and {Ackermann}, M. and {Adams}, J. and {Aguilar}, J.~A. and {Ahlers}, M. and {Ahrens}, M. and {Samarai}, I. Al and {Altmann}, D. and {Andeen}, K. and {Anderson}, T. and {Ansseau}, I. and {Anton}, G. and {Arg{\"u}elles}, C. and {Arsioli}, B. and {Auffenberg}, J. and {Axani}, S. and {Bagherpour}, H. and {Bai}, X. and {Barron}, J.~P. and {Barwick}, S.~W. and {Baum}, V. and {Bay}, R. and {Beatty}, J.~J. and {Becker Tjus}, J. and {Becker}, K. -H. and {BenZvi}, S. and {Berley}, D. and {Bernardini}, E. and {Besson}, D.~Z. and {Binder}, G. and {Bindig}, D. and {Blaufuss}, E. and {Blot}, S. and {Bohm}, C. and {B{\"o}rner}, M. and {Bos}, F. and {B{\"o}ser}, S. and {Botner}, O. and {Bourbeau}, E. and {Bourbeau}, J. and {Bradascio}, F. and {Braun}, J. and {Brenzke}, M. and {Bretz}, H. -P. and {Bron}, S. and {Brostean-Kaiser}, J. and {Burgman}, A. and {Busse}, R.~S. and {Carver}, T. and {Cheung}, E. and {Chirkin}, D. and {Christov}, A. and {Clark}, K. and {Classen}, L. and {Coenders}, S. and {Collin}, G.~H. and {Conrad}, J.~M. and {Coppin}, P. and {Correa}, P. and {Cowen}, D.~F. and {Cross}, R. and {Dave}, P. and {Day}, M. and {de Andr{\'e}}, J.~P.~A.~M. and {De Clercq}, C. and {DeLaunay}, J.~J. and {Dembinski}, H. and {DeRidder}, S. and {Desiati}, P. and {de Vries}, K.~D. and {de Wasseige}, G. and {de With}, M. and {DeYoung}, T. and {D{\'\i}az-V{\'e}lez}, J.~C. and {di Lorenzo}, V. and {Dujmovic}, H. and {Dumm}, J.~P. and {Dunkman}, M. and {Dvorak}, E. and {Eberhardt}, B. and {Ehrhardt}, T. and {Eichmann}, B. and {Eller}, P. and {Evenson}, P.~A. and {Fahey}, S. and {Fazely}, A.~R. and {Felde}, J. and {Filimonov}, K. and {Finley}, C. and {Flis}, S. and {Franckowiak}, A. and {Friedman}, E. and {Fritz}, A. and {Gaisser}, T.~K. and {Gallagher}, J. and {Gerhardt}, L. and {Ghorbani}, K. and {Giommi}, P. and {Glauch}, T. and {Gl{\"u}senkamp}, T. and {Goldschmidt}, A. and {Gonzalez}, J.~G. and {Grant}, D. and {Griffith}, Z. and {Haack}, C. and {Hallgren}, A. and {Halzen}, F. and {Hanson}, K. and {Hebecker}, D. and {Heereman}, D. and {Helbing}, K. and {Hellauer}, R. and {Hickford}, S. and {Hignight}, J. and {Hill}, G.~C. and {Hoffman}, K.~D. and {Hoffmann}, R. and {Hoinka}, T. and {Hokanson-Fasig}, B. and {Hoshina}, K. and {Huang}, F. and {Huber}, M. and {Hultqvist}, K. and {H{\"u}nnefeld}, M. and {Hussain}, R. and {In}, S. and {Iovine}, N. and {Ishihara}, A. and {Jacobi}, E. and {Japaridze}, G.~S. and {Jeong}, M. and {Jero}, K. and {Jones}, B.~J.~P. and {Kalaczynski}, P. and {Kang}, W. and {Kappes}, A. and {Kappesser}, D. and {Karg}, T. and {Karle}, A. and {Katz}, U. and {Kauer}, M. and {Keivani}, A. and {Kelley}, J.~L. and {Kheirandish}, A. and {Kim}, J. and {Kim}, M. and {Kintscher}, T. and {Kiryluk}, J. and {Kittler}, T. and {Klein}, S.~R. and {Koirala}, R. and {Kolanoski}, H. and {K{\"o}pke}, L. and {Kopper}, C. and {Kopper}, S. and {Koschinsky}, J.~P. and {Koskinen}, D.~J. and {Kowalski}, M. and {Krammer}, B. and {Krings}, K. and {Kroll}, M. and {Kr{\"u}ckl}, G. and {Kunwar}, S. and {Kurahashi}, N. and {Kuwabara}, T. and {Kyriacou}, A. and {Labare}, M. and {Lanfranchi}, J.~L. and {Larson}, M.~J. and {Lauber}, F. and {Leonard}, K. and {Lesiak-Bzdak}, M. and {Leuermann}, M. and {Liu}, Q.~R. and {Lozano Mariscal}, C.~J. and {Lu}, L. and {L{\"u}nemann}, J. and {Luszczak}, W. and {Madsen}, J. and {Maggi}, G. and {Mahn}, K.~B.~M. and {Mancina}, S. and {Maruyama}, R. and {Mase}, K. and {Maunu}, R. and {Meagher}, K. and {Medici}, M. and {Meier}, M. and {Menne}, T. and {Merino}, G. and {Meures}, T. and {Miarecki}, S. and {Micallef}, J. and {Moment{\'e}}, G. and {Montaruli}, T. and {Moore}, R.~W. and {Morse}, R. and {Moulai}, M. and {Nahnhauer}, R.},
        title = "{Neutrino emission from the direction of the blazar TXS 0506+056 prior to the IceCube-170922A alert}",
      journal = {Science},
         year = 2018,
        month = jul,
       volume = {361},
       number = {6398},
        pages = {147-151},
          doi = {10.1126/science.aat2890},
archivePrefix = {arXiv},
       eprint = {1807.08794},
 primaryClass = {astro-ph.HE},
       adsurl = {https://ui.adsabs.harvard.edu/abs/2018Sci...361..147I}
}

@ARTICLE{IceCube2018b-TXS,
       author = {{IceCube Collaboration} and {Aartsen}, M.~G. and {Ackermann}, M. and {Adams}, J. and {Aguilar}, J.~A. and {Ahlers}, M. and {Ahrens}, M. and {Al Samarai}, I. and {Altmann}, D. and {Andeen}, K. and {Anderson}, T. and {Ansseau}, I. and {Anton}, G. and {Arg{\"u}elles}, C. and {Auffenberg}, J. and {Axani}, S. and {Bagherpour}, H. and {Bai}, X. and {Barron}, J.~P. and {Barwick}, S.~W. and {Baum}, V. and {Bay}, R. and {Beatty}, J.~J. and {Becker Tjus}, J. and {Becker}, K. -H. and {BenZvi}, S. and {Berley}, D. and {Bernardini}, E. and {Besson}, D.~Z. and {Binder}, G. and {Bindig}, D. and {Blaufuss}, E. and {Blot}, S. and {Bohm}, C. and {B{\"o}rner}, M. and {Bos}, F. and {B{\"o}ser}, S. and {Botner}, O. and {Bourbeau}, E. and {Bourbeau}, J. and {Bradascio}, F. and {Braun}, J. and {Brenzke}, M. and {Bretz}, H. -P. and {Bron}, S. and {Brostean-Kaiser}, J. and {Burgman}, A. and {Busse}, R.~S. and {Carver}, T. and {Cheung}, E. and {Chirkin}, D. and {Christov}, A. and {Clark}, K. and {Classen}, L. and {Coenders}, S. and {Collin}, G.~H. and {Conrad}, J.~M. and {Coppin}, P. and {Correa}, P. and {Cowen}, D.~F. and {Cross}, R. and {Dave}, P. and {Day}, M. and {de Andr{\'e}}, J.~P.~A.~M. and {De Clercq}, C. and {DeLaunay}, J.~J. and {Dembinski}, H. and {De Ridder}, S. and {Desiati}, P. and {de Vries}, K.~D. and {de Wasseige}, G. and {de With}, M. and {DeYoung}, T. and {D{\'\i}az-V{\'e}lez}, J.~C. and {di Lorenzo}, V. and {Dujmovic}, H. and {Dumm}, J.~P. and {Dunkman}, M. and {Dvorak}, E. and {Eberhardt}, B. and {Ehrhardt}, T. and {Eichmann}, B. and {Eller}, P. and {Evenson}, P.~A. and {Fahey}, S. and {Fazely}, A.~R. and {Felde}, J. and {Filimonov}, K. and {Finley}, C. and {Flis}, S. and {Franckowiak}, A. and {Friedman}, E. and {Fritz}, A. and {Gaisser}, T.~K. and {Gallagher}, J. and {Gerhardt}, L. and {Ghorbani}, K. and {Glauch}, T. and {Gl{\"u}senkamp}, T. and {Goldschmidt}, A. and {Gonzalez}, J.~G. and {Grant}, D. and {Griffith}, Z. and {Haack}, C. and {Hallgren}, A. and {Halzen}, F. and {Hanson}, K. and {Hebecker}, D. and {Heereman}, D. and {Helbing}, K. and {Hellauer}, R. and {Hickford}, S. and {Hignight}, J. and {Hill}, G.~C. and {Hoffman}, K.~D. and {Hoffmann}, R. and {Hoinka}, T. and {Hokanson-Fasig}, B. and {Hoshina}, K. and {Huang}, F. and {Huber}, M. and {Hultqvist}, K. and {H{\"u}nnefeld}, M. and {Hussain}, R. and {In}, S. and {Iovine}, N. and {Ishihara}, A. and {Jacobi}, E. and {Japaridze}, G.~S. and {Jeong}, M. and {Jero}, K. and {Jones}, B.~J.~P. and {Kalaczynski}, P. and {Kang}, W. and {Kappes}, A. and {Kappesser}, D. and {Karg}, T. and {Karle}, A. and {Katz}, U. and {Kauer}, M. and {Keivani}, A. and {Kelley}, J.~L. and {Kheirandish}, A. and {Kim}, J. and {Kim}, M. and {Kintscher}, T. and {Kiryluk}, J. and {Kittler}, T. and {Klein}, S.~R. and {Koirala}, R. and {Kolanoski}, H. and {K{\"o}pke}, L. and {Kopper}, C. and {Kopper}, S. and {Koschinsky}, J.~P. and {Koskinen}, D.~J. and {Kowalski}, M. and {Krings}, K. and {Kroll}, M. and {Kr{\"u}ckl}, G. and {Kunwar}, S. and {Kurahashi}, N. and {Kuwabara}, T. and {Kyriacou}, A. and {Labare}, M. and {Lanfranchi}, J.~L. and {Larson}, M.~J. and {Lauber}, F. and {Leonard}, K. and {Lesiak-Bzdak}, M. and {Leuermann}, M. and {Liu}, Q.~R. and {Lozano Mariscal}, C.~J. and {Lu}, L. and {L{\"u}nemann}, J. and {Luszczak}, W. and {Madsen}, J. and {Maggi}, G. and {Mahn}, K.~B.~M. and {Mancina}, S. and {Maruyama}, R. and {Mase}, K. and {Maunu}, R. and {Meagher}, K. and {Medici}, M. and {Meier}, M. and {Menne}, T. and {Merino}, G. and {Meures}, T. and {Miarecki}, S. and {Micallef}, J. and {Moment{\'e}}, G. and {Montaruli}, T. and {Moore}, R.~W. and {Morse}, R. and {Moulai}, M. and {Nahnhauer}, R. and {Nakarmi}, P. and {Naumann}, U. and {Neer}, G.},
        title = "{Multimessenger observations of a flaring blazar coincident with high-energy neutrino IceCube-170922A}",
      journal = {Science},
         year = 2018,
        month = jul,
       volume = {361},
       number = {6398},
          eid = {eaat1378},
        pages = {eaat1378},
          doi = {10.1126/science.aat1378},
archivePrefix = {arXiv},
       eprint = {1807.08816},
 primaryClass = {astro-ph.HE},
       adsurl = {https://ui.adsabs.harvard.edu/abs/2018Sci...361.1378I}
}

@ARTICLE{Stein2021-TDE-dsg,
       author = {{Stein}, Robert and {van Velzen}, Sjoert and {Kowalski}, Marek and {Franckowiak}, Anna and {Gezari}, Suvi and {Miller-Jones}, James C.~A. and {Frederick}, Sara and {Sfaradi}, Itai and {Bietenholz}, Michael F. and {Horesh}, Assaf and {Fender}, Rob and {Garrappa}, Simone and {Ahumada}, Tom{\'a}s and {Andreoni}, Igor and {Belicki}, Justin and {Bellm}, Eric C. and {B{\"o}ttcher}, Markus and {Brinnel}, Valery and {Burruss}, Rick and {Cenko}, S. Bradley and {Coughlin}, Michael W. and {Cunningham}, Virginia and {Drake}, Andrew and {Farrar}, Glennys R. and {Feeney}, Michael and {Foley}, Ryan J. and {Gal-Yam}, Avishay and {Golkhou}, V. Zach and {Goobar}, Ariel and {Graham}, Matthew J. and {Hammerstein}, Erica and {Helou}, George and {Hung}, Tiara and {Kasliwal}, Mansi M. and {Kilpatrick}, Charles D. and {Kong}, Albert K.~H. and {Kupfer}, Thomas and {Laher}, Russ R. and {Mahabal}, Ashish A. and {Masci}, Frank J. and {Necker}, Jannis and {Nordin}, Jakob and {Perley}, Daniel A. and {Rigault}, Mickael and {Reusch}, Simeon and {Rodriguez}, Hector and {Rojas-Bravo}, C{\'e}sar and {Rusholme}, Ben and {Shupe}, David L. and {Singer}, Leo P. and {Sollerman}, Jesper and {Soumagnac}, Maayane T. and {Stern}, Daniel and {Taggart}, Kirsty and {van Santen}, Jakob and {Ward}, Charlotte and {Woudt}, Patrick and {Yao}, Yuhan},
        title = "{A tidal disruption event coincident with a high-energy neutrino}",
      journal = {Nature Astronomy},
         year = 2021,
        month = feb,
       volume = {5},
        pages = {510-518},
          doi = {10.1038/s41550-020-01295-8},
archivePrefix = {arXiv},
       eprint = {2005.05340},
 primaryClass = {astro-ph.HE},
       adsurl = {https://ui.adsabs.harvard.edu/abs/2021NatAs...5..510S}
}

@ARTICLE{Reusch2022-TDE-fdr,
       author = {{Reusch}, Simeon and {Stein}, Robert and {Kowalski}, Marek and {van Velzen}, Sjoert and {Franckowiak}, Anna and {Lunardini}, Cecilia and {Murase}, Kohta and {Winter}, Walter and {Miller-Jones}, James C.~A. and {Kasliwal}, Mansi M. and {Gilfanov}, Marat and {Garrappa}, Simone and {Paliya}, Vaidehi S. and {Ahumada}, Tom{\'a}s and {Anand}, Shreya and {Barbarino}, Cristina and {Bellm}, Eric C. and {Brinnel}, Val{\'e}ry and {Buson}, Sara and {Cenko}, S. Bradley and {Coughlin}, Michael W. and {De}, Kishalay and {Dekany}, Richard and {Frederick}, Sara and {Gal-Yam}, Avishay and {Gezari}, Suvi and {Giroletti}, Marcello and {Graham}, Matthew J. and {Karambelkar}, Viraj and {Kimura}, Shigeo S. and {Kong}, Albert K.~H. and {Kool}, Erik C. and {Laher}, Russ R. and {Medvedev}, Pavel and {Necker}, Jannis and {Nordin}, Jakob and {Perley}, Daniel A. and {Rigault}, Mickael and {Rusholme}, Ben and {Schulze}, Steve and {Schweyer}, Tassilo and {Singer}, Leo P. and {Sollerman}, Jesper and {Strotjohann}, Nora Linn and {Sunyaev}, Rashid and {van Santen}, Jakob and {Walters}, Richard and {Zhang}, B. Theodore and {Zimmerman}, Erez},
        title = "{Candidate Tidal Disruption Event AT2019fdr Coincident with a High-Energy Neutrino}",
      journal = {\prl},
         year = 2022,
        month = jun,
       volume = {128},
       number = {22},
          eid = {221101},
        pages = {221101},
          doi = {10.1103/PhysRevLett.128.221101},
archivePrefix = {arXiv},
       eprint = {2111.09390},
 primaryClass = {astro-ph.HE},
       adsurl = {https://ui.adsabs.harvard.edu/abs/2022PhRvL.128v1101R}
}

@ARTICLE{van_Velzen2024-TDE-aalc,
       author = {{van Velzen}, Sjoert and {Stein}, Robert and {Gilfanov}, Marat and {Kowalski}, Marek and {Hayasaki}, Kimitake and {Reusch}, Simeon and {Yao}, Yuhan and {Garrappa}, Simone and {Franckowiak}, Anna and {Gezari}, Suvi and {Nordin}, Jakob and {Fremling}, Christoffer and {Sharma}, Yashvi and {Yan}, Lin and {Kool}, Erik C. and {Stern}, Daniel and {Veres}, Patrik M. and {Sollerman}, Jesper and {Medvedev}, Pavel and {Sunyaev}, Rashid and {Bellm}, Eric C. and {Dekany}, Richard G. and {Duev}, Dimitri A. and {Graham}, Matthew J. and {Kasliwal}, Mansi M. and {Kulkarni}, Shrinivas R. and {Laher}, Russ R. and {Riddle}, Reed L. and {Rusholme}, Ben},
        title = "{Establishing accretion flares from supermassive black holes as a source of high-energy neutrinos}",
      journal = {\mnras},
         year = 2024,
        month = apr,
       volume = {529},
       number = {3},
        pages = {2559-2576},
          doi = {10.1093/mnras/stae610},
archivePrefix = {arXiv},
       eprint = {2111.09391},
 primaryClass = {astro-ph.HE},
       adsurl = {https://ui.adsabs.harvard.edu/abs/2024MNRAS.529.2559V}
}

@ARTICLE{IceCube2022-1068,
       author = {{IceCube Collaboration} and {Abbasi}, R. and {Ackermann}, M. and {Adams}, J. and {Aguilar}, J.~A. and {Ahlers}, M. and {Ahrens}, M. and {Alameddine}, J.~M. and {Alispach}, C. and {Alves}, Jr., A.~A. and {Amin}, N.~M. and {Andeen}, K. and {Anderson}, T. and {Anton}, G. and {Arg{\"u}elles}, C. and {Ashida}, Y. and {Axani}, S. and {Bai}, X. and {Balagopal}, A.~V. and {Barbano}, V.~A. and {Barwick}, S.~W. and {Bastian}, B. and {Basu}, V. and {Baur}, S. and {Bay}, R. and {Beatty}, J.~J. and {Becker}, K. -H. and {Becker Tjus}, J. and {Bellenghi}, C. and {Benzvi}, S. and {Berley}, D. and {Bernardini}, E. and {Besson}, D.~Z. and {Binder}, G. and {Bindig}, D. and {Blaufuss}, E. and {Blot}, S. and {Boddenberg}, M. and {Bontempo}, F. and {Borowka}, J. and {B{\"o}ser}, S. and {Botner}, O. and {B{\"o}ttcher}, J. and {Bourbeau}, E. and {Bradascio}, F. and {Braun}, J. and {Brinson}, B. and {Bron}, S. and {Brostean-Kaiser}, J. and {Browne}, S. and {Burgman}, A. and {Burley}, R.~T. and {Busse}, R.~S. and {Campana}, M.~A. and {Carnie-Bronca}, E.~G. and {Chen}, C. and {Chen}, Z. and {Chirkin}, D. and {Choi}, K. and {Clark}, B.~A. and {Clark}, K. and {Classen}, L. and {Coleman}, A. and {Collin}, G.~H. and {Conrad}, J.~M. and {Coppin}, P. and {Correa}, P. and {Cowen}, D.~F. and {Cross}, R. and {Dappen}, C. and {Dave}, P. and {de Clercq}, C. and {Delaunay}, J.~J. and {Delgado L{\'o}pez}, D. and {Dembinski}, H. and {Deoskar}, K. and {Desai}, A. and {Desiati}, P. and {de Vries}, K.~D. and {de Wasseige}, G. and {de With}, M. and {Deyoung}, T. and {Diaz}, A. and {D{\'\i}az-V{\'e}lez}, J.~C. and {Dittmer}, M. and {Dujmovic}, H. and {Dunkman}, M. and {Duvernois}, M.~A. and {Dvorak}, E. and {Ehrhardt}, T. and {Eller}, P. and {Engel}, R. and {Erpenbeck}, H. and {Evans}, J. and {Evenson}, P.~A. and {Fan}, K.~L. and {Fazely}, A.~R. and {Fedynitch}, A. and {Feigl}, N. and {Fiedlschuster}, S. and {Fienberg}, A.~T. and {Filimonov}, K. and {Finley}, C. and {Fischer}, L. and {Fox}, D. and {Franckowiak}, A. and {Friedman}, E. and {Fritz}, A. and {F{\"u}rst}, P. and {Gaisser}, T.~K. and {Gallagher}, J. and {Ganster}, E. and {Garcia}, A. and {Garrappa}, S. and {Gerhardt}, L. and {Ghadimi}, A. and {Glaser}, C. and {Glauch}, T. and {Gl{\"u}senkamp}, T. and {Goldschmidt}, A. and {Gonzalez}, J.~G. and {Goswami}, S. and {Grant}, D. and {Gr{\'e}goire}, T. and {Griswold}, S. and {G{\"u}nther}, C. and {Gutjahr}, P. and {Haack}, C. and {Hallgren}, A. and {Halliday}, R. and {Halve}, L. and {Halzen}, F. and {Hanson}, M. Ha Minh K. and {Hardin}, J. and {Harnisch}, A.~A. and {Haungs}, A. and {Hebecker}, D. and {Helbing}, K. and {Henningsen}, F. and {Hettinger}, E.~C. and {Hickford}, S. and {Hignight}, J. and {Hill}, C. and {Hill}, G.~C. and {Hoffman}, K.~D. and {Hoffmann}, R. and {Hokanson-Fasig}, B. and {Hoshina}, K. and {Huang}, F. and {Huber}, M. and {Huber}, T. and {Hultqvist}, K. and {H{\"u}nnefeld}, M. and {Hussain}, R. and {Hymon}, K. and {in}, S. and {Iovine}, N. and {Ishihara}, A. and {Jansson}, M. and {Japaridze}, G.~S. and {Jeong}, M. and {Jin}, M. and {Jones}, B.~J.~P. and {Kang}, D. and {Kang}, W. and {Kang}, X. and {Kappes}, A. and {Kappesser}, D. and {Kardum}, L. and {Karg}, T. and {Karl}, M. and {Karle}, A. and {Katz}, U. and {Kauer}, M. and {Kellermann}, M. and {Kelley}, J.~L. and {Kheirandish}, A. and {Kin}, K. and {Kintscher}, T. and {Kiryluk}, J. and {Klein}, S.~R. and {Koirala}, R. and {Kolanoski}, H. and {Kontrimas}, T. and {K{\"o}pke}, L. and {Kopper}, C. and {Kopper}, S. and {Koskinen}, D.~J. and {Koundal}, P. and {Kovacevich}, M. and {Kowalski}, M. and {Kozynets}, T. and {Kun}, E. and {Kurahashi}, N. and {Lad}, N. and {Lagunas Gualda}, C. and {Lanfranchi}, J.~L. and {Larson}, M.~J. and {Lauber}, F. and {Lazar}, J.~P.},
        title = "{Evidence for neutrino emission from the nearby active galaxy NGC 1068}",
      journal = {Science},
         year = 2022,
        month = nov,
       volume = {378},
       number = {6619},
        pages = {538-543},
          doi = {10.1126/science.abg3395},
archivePrefix = {arXiv},
       eprint = {2211.09972},
 primaryClass = {astro-ph.HE},
       adsurl = {https://ui.adsabs.harvard.edu/abs/2022Sci...378..538I}
}

@ARTICLE{Abbasi2025-1068-4151,
       author = {{Abbasi}, R. and {Ackermann}, M. and {Adams}, J. and {Agarwalla}, S.~K. and {Aguilar}, J.~A. and {Ahlers}, M. and {Alameddine}, J.~M. and {Amin}, N.~M. and {Andeen}, K. and {Arg{\"u}elles}, C. and {Ashida}, Y. and {Athanasiadou}, S. and {Ausborm}, L. and {Axani}, S.~N. and {Bai}, X. and {Balagopal V.}, A. and {Baricevic}, M. and {Barwick}, S.~W. and {Bash}, S. and {Basu}, V. and {Bay}, R. and {Beatty}, J.~J. and {Becker Tjus}, J. and {Beise}, J. and {Bellenghi}, C. and {Benning}, C. and {BenZvi}, S. and {Berley}, D. and {Bernardini}, E. and {Besson}, D.~Z. and {Blaufuss}, E. and {Bloom}, L. and {Blot}, S. and {Bontempo}, F. and {Book Motzkin}, J.~Y. and {Boscolo Meneguolo}, C. and {B{\"o}ser}, S. and {Botner}, O. and {B{\"o}ttcher}, J. and {Braun}, J. and {Brinson}, B. and {Brostean-Kaiser}, J. and {Brusa}, L. and {Burley}, R.~T. and {Butterfield}, D. and {Campana}, M.~A. and {Caracas}, I. and {Carloni}, K. and {Carpio}, J. and {Chattopadhyay}, S. and {Chau}, N. and {Chen}, Z. and {Chirkin}, D. and {Choi}, S. and {Clark}, B.~A. and {Coleman}, A. and {Collin}, G.~H. and {Connolly}, A. and {Conrad}, J.~M. and {Coppin}, P. and {Corley}, R. and {Correa}, P. and {Cowen}, D.~F. and {Dave}, P. and {De Clercq}, C. and {DeLaunay}, J.~J. and {Delgado}, D. and {Deng}, S. and {Desai}, A. and {Desiati}, P. and {de Vries}, K.~D. and {de Wasseige}, G. and {DeYoung}, T. and {Diaz}, A. and {D{\'\i}az-V{\'e}lez}, J.~C. and {Dierichs}, P. and {Dittmer}, M. and {Domi}, A. and {Draper}, L. and {Dujmovic}, H. and {Dutta}, K. and {DuVernois}, M.~A. and {Ehrhardt}, T. and {Eidenschink}, L. and {Eimer}, A. and {Eller}, P. and {Ellinger}, E. and {El Mentawi}, S. and {Els{\"a}sser}, D. and {Engel}, R. and {Erpenbeck}, H. and {Evans}, J. and {Evenson}, P.~A. and {Fan}, K.~L. and {Fang}, K. and {Farrag}, K. and {Fazely}, A.~R. and {Fedynitch}, A. and {Feigl}, N. and {Fiedlschuster}, S. and {Finley}, C. and {Fischer}, L. and {Fox}, D. and {Franckowiak}, A. and {Fukami}, S. and {F{\"u}rst}, P. and {Gallagher}, J. and {Ganster}, E. and {Garcia}, A. and {Garcia}, M. and {Garg}, G. and {Genton}, E. and {Gerhardt}, L. and {Ghadimi}, A. and {Girard-Carillo}, C. and {Glaser}, C. and {Gl{\"u}senkamp}, T. and {Gonzalez}, J.~G. and {Goswami}, S. and {Granados}, A. and {Grant}, D. and {Gray}, S.~J. and {Gries}, O. and {Griffin}, S. and {Griswold}, S. and {Groth}, K.~M. and {G{\"u}nther}, C. and {Gutjahr}, P. and {Ha}, C. and {Haack}, C. and {Hallgren}, A. and {Halve}, L. and {Halzen}, F. and {Hamdaoui}, H. and {Ha Minh}, M. and {Handt}, M. and {Hanson}, K. and {Hardin}, J. and {Harnisch}, A.~A. and {Hatch}, P. and {Haungs}, A. and {H{\"a}ussler}, J. and {Helbing}, K. and {Hellrung}, J. and {Hermannsgabner}, J. and {Heuermann}, L. and {Heyer}, N. and {Hickford}, S. and {Hidvegi}, A. and {Hill}, C. and {Hill}, G.~C. and {Hoffman}, K.~D. and {Hori}, S. and {Hoshina}, K. and {Hostert}, M. and {Hou}, W. and {Huber}, T. and {Hultqvist}, K. and {H{\"u}nnefeld}, M. and {Hussain}, R. and {Hymon}, K. and {Ishihara}, A. and {Iwakiri}, W. and {Jacquart}, M. and {Janik}, O. and {Jansson}, M. and {Japaridze}, G.~S. and {Jeong}, M. and {Jin}, M. and {Jones}, B.~J.~P. and {Kamp}, N. and {Kang}, D. and {Kang}, W. and {Kang}, X. and {Kappes}, A. and {Kappesser}, D. and {Kardum}, L. and {Karg}, T. and {Karl}, M. and {Karle}, A. and {Katil}, A. and {Katz}, U. and {Kauer}, M. and {Kelley}, J.~L. and {Khanal}, M. and {Khatee Zathul}, A. and {Kheirandish}, A. and {Kiryluk}, J. and {Klein}, S.~R. and {Kochocki}, A. and {Koirala}, R. and {Kolanoski}, H. and {Kontrimas}, T. and {K{\"o}pke}, L. and {Kopper}, C. and {Koskinen}, D.~J. and {Koundal}, P. and {Kovacevich}, M. and {Kowalski}, M. and {Kozynets}, T.},
        title = "{Search for Neutrino Emission from Hard X-Ray AGN with IceCube}",
      journal = {\apj},
         year = 2025,
        month = mar,
       volume = {981},
       number = {2},
          eid = {131},
        pages = {131},
          doi = {10.3847/1538-4357/ada94b},
archivePrefix = {arXiv},
       eprint = {2406.06684},
 primaryClass = {astro-ph.HE},
       adsurl = {https://ui.adsabs.harvard.edu/abs/2025ApJ...981..131A}
}

@ARTICLE{Stecker1991,
       author = {{Stecker}, F.~W. and {Done}, C. and {Salamon}, M.~H. and {Sommers}, P.},
        title = "{High-energy neutrinos from active galactic nuclei}",
      journal = {\prl},
         year = 1991,
        month = may,
       volume = {66},
       number = {21},
        pages = {2697-2700},
          doi = {10.1103/PhysRevLett.66.2697},
       adsurl = {https://ui.adsabs.harvard.edu/abs/1991PhRvL..66.2697S}
}

@ARTICLE{Halzen1997,
       author = {{Halzen}, F. and {Zas}, E.},
        title = "{Neutrino Fluxes from Active Galaxies: A Model-Independent Estimate}",
      journal = {\apj},
         year = 1997,
        month = oct,
       volume = {488},
       number = {2},
        pages = {669-674},
          doi = {10.1086/304741},
archivePrefix = {arXiv},
       eprint = {astro-ph/9702193},
 primaryClass = {astro-ph},
       adsurl = {https://ui.adsabs.harvard.edu/abs/1997ApJ...488..669H}
}

@ARTICLE{Aartsen2020-10yr,
       author = {{Aartsen}, M.~G. and {Ackermann}, M. and {Adams}, J. and {Aguilar}, J.~A. and {Ahlers}, M. and {Ahrens}, M. and {Alispach}, C. and {Andeen}, K. and {Anderson}, T. and {Ansseau}, I. and {Anton}, G. and {Arg{\"u}elles}, C. and {Auffenberg}, J. and {Axani}, S. and {Backes}, P. and {Bagherpour}, H. and {Bai}, X. and {Balagopal}, A. and {Barbano}, A. and {Barwick}, S.~W. and {Bastian}, B. and {Baum}, V. and {Baur}, S. and {Bay}, R. and {Beatty}, J.~J. and {Becker}, K. -H. and {Becker Tjus}, J. and {BenZvi}, S. and {Berley}, D. and {Bernardini}, E. and {Besson}, D.~Z. and {Binder}, G. and {Bindig}, D. and {Blaufuss}, E. and {Blot}, S. and {Bohm}, C. and {B{\"o}rner}, M. and {B{\"o}ser}, S. and {Botner}, O. and {B{\"o}ttcher}, J. and {Bourbeau}, E. and {Bourbeau}, J. and {Bradascio}, F. and {Braun}, J. and {Bron}, S. and {Brostean-Kaiser}, J. and {Burgman}, A. and {Buscher}, J. and {Busse}, R.~S. and {Carver}, T. and {Chen}, C. and {Cheung}, E. and {Chirkin}, D. and {Choi}, S. and {Clark}, K. and {Classen}, L. and {Coleman}, A. and {Collin}, G.~H. and {Conrad}, J.~M. and {Coppin}, P. and {Correa}, P. and {Cowen}, D.~F. and {Cross}, R. and {Dave}, P. and {De Clercq}, C. and {DeLaunay}, J.~J. and {Dembinski}, H. and {Deoskar}, K. and {De Ridder}, S. and {Desiati}, P. and {de Vries}, K.~D. and {de Wasseige}, G. and {de With}, M. and {DeYoung}, T. and {Diaz}, A. and {D{\'\i}az-V{\'e}lez}, J.~C. and {Dujmovic}, H. and {Dunkman}, M. and {Dvorak}, E. and {Eberhardt}, B. and {Ehrhardt}, T. and {Eller}, P. and {Engel}, R. and {Evenson}, P.~A. and {Fahey}, S. and {Fazely}, A.~R. and {Felde}, J. and {Filimonov}, K. and {Finley}, C. and {Fox}, D. and {Franckowiak}, A. and {Friedman}, E. and {Fritz}, A. and {Gaisser}, T.~K. and {Gallagher}, J. and {Ganster}, E. and {Garrappa}, S. and {Gerhardt}, L. and {Ghorbani}, K. and {Glauch}, T. and {Gl{\"u}senkamp}, T. and {Goldschmidt}, A. and {Gonzalez}, J.~G. and {Grant}, D. and {Griffith}, Z. and {Griswold}, S. and {G{\"u}nder}, M. and {G{\"u}nd{\"u}z}, M. and {Haack}, C. and {Hallgren}, A. and {Halliday}, R. and {Halve}, L. and {Halzen}, F. and {Hanson}, K. and {Haungs}, A. and {Hebecker}, D. and {Heereman}, D. and {Heix}, P. and {Helbing}, K. and {Hellauer}, R. and {Henningsen}, F. and {Hickford}, S. and {Hignight}, J. and {Hill}, G.~C. and {Hoffman}, K.~D. and {Hoffmann}, R. and {Hoinka}, T. and {Hokanson-Fasig}, B. and {Hoshina}, K. and {Huang}, F. and {Huber}, M. and {Huber}, T. and {Hultqvist}, K. and {H{\"u}nnefeld}, M. and {Hussain}, R. and {In}, S. and {Iovine}, N. and {Ishihara}, A. and {Japaridze}, G.~S. and {Jeong}, M. and {Jero}, K. and {Jones}, B.~J.~P. and {Jonske}, F. and {Joppe}, R. and {Kang}, D. and {Kang}, W. and {Kappes}, A. and {Kappesser}, D. and {Karg}, T. and {Karl}, M. and {Karle}, A. and {Katz}, U. and {Kauer}, M. and {Kelley}, J.~L. and {Kheirandish}, A. and {Kim}, J. and {Kintscher}, T. and {Kiryluk}, J. and {Kittler}, T. and {Klein}, S.~R. and {Koirala}, R. and {Kolanoski}, H. and {K{\"o}pke}, L. and {Kopper}, C. and {Kopper}, S. and {Koskinen}, D.~J. and {Kowalski}, M. and {Krings}, K. and {Kr{\"u}ckl}, G. and {Kulacz}, N. and {Kurahashi}, N. and {Kyriacou}, A. and {Labare}, M. and {Lanfranchi}, J.~L. and {Larson}, M.~J. and {Lauber}, F. and {Lazar}, J.~P. and {Leonard}, K. and {Leszczy{\'n}ska}, A. and {Leuermann}, M. and {Liu}, Q.~R. and {Lohfink}, E. and {Lozano Mariscal}, C.~J. and {Lu}, L. and {Lucarelli}, F. and {L{\"u}nemann}, J. and {Luszczak}, W. and {Lyu}, Y. and {Ma}, W.~Y. and {Madsen}, J. and {Maggi}, G. and {Mahn}, K.~B.~M. and {Makino}, Y. and {Mallik}, P. and {Mallot}, K. and {Mancina}, S. and {Mari{\c{s}}}, I.~C. and {Maruyama}, R. and {Mase}, K. and {Matis}, H.~S.},
        title = "{Time-Integrated Neutrino Source Searches with 10 Years of IceCube Data}",
      journal = {\prl},
         year = 2020,
        month = feb,
       volume = {124},
       number = {5},
          eid = {051103},
        pages = {051103},
          doi = {10.1103/PhysRevLett.124.051103},
archivePrefix = {arXiv},
       eprint = {1910.08488},
 primaryClass = {astro-ph.HE},
       adsurl = {https://ui.adsabs.harvard.edu/abs/2020PhRvL.124e1103A}
}

@ARTICLE{Hwang2020,
       author = {{Hwang}, Hsiang-Chih and {Zakamska}, Nadia L.},
        title = "{Lifetime of short-period binaries measured from their Galactic kinematics}",
      journal = {\mnras},
         year = 2020,
        month = apr,
       volume = {493},
       number = {2},
        pages = {2271-2286},
          doi = {10.1093/mnras/staa400},
archivePrefix = {arXiv},
       eprint = {1909.06375},
 primaryClass = {astro-ph.SR},
       adsurl = {https://ui.adsabs.harvard.edu/abs/2020MNRAS.493.2271H}
}

@ARTICLE{Keivani2018-blazar-jet,
       author = {{Keivani}, A. and {Murase}, K. and {Petropoulou}, M. and {Fox}, D.~B. and {Cenko}, S.~B. and {Chaty}, S. and {Coleiro}, A. and {DeLaunay}, J.~J. and {Dimitrakoudis}, S. and {Evans}, P.~A. and {Kennea}, J.~A. and {Marshall}, F.~E. and {Mastichiadis}, A. and {Osborne}, J.~P. and {Santander}, M. and {Tohuvavohu}, A. and {Turley}, C.~F.},
        title = "{A Multimessenger Picture of the Flaring Blazar TXS 0506+056: Implications for High-energy Neutrino Emission and Cosmic-Ray Acceleration}",
      journal = {\apj},
         year = 2018,
        month = sep,
       volume = {864},
       number = {1},
          eid = {84},
        pages = {84},
          doi = {10.3847/1538-4357/aad59a},
archivePrefix = {arXiv},
       eprint = {1807.04537},
 primaryClass = {astro-ph.HE},
       adsurl = {https://ui.adsabs.harvard.edu/abs/2018ApJ...864...84K}
}

@ARTICLE{Cerruti2019-blazar-jet,
       author = {{Cerruti}, M. and {Zech}, A. and {Boisson}, C. and {Emery}, G. and {Inoue}, S. and {Lenain}, J. -P.},
        title = "{Leptohadronic single-zone models for the electromagnetic and neutrino emission of TXS 0506+056}",
      journal = {\mnras},
         year = 2019,
        month = feb,
       volume = {483},
       number = {1},
        pages = {L12-L16},
          doi = {10.1093/mnrasl/sly210},
archivePrefix = {arXiv},
       eprint = {1807.04335},
 primaryClass = {astro-ph.HE},
       adsurl = {https://ui.adsabs.harvard.edu/abs/2019MNRAS.483L..12C}
}

@ARTICLE{Gao2019-blazar-jet,
       author = {{Gao}, Shan and {Fedynitch}, Anatoli and {Winter}, Walter and {Pohl}, Martin},
        title = "{Modelling the coincident observation of a high-energy neutrino and a bright blazar flare}",
      journal = {Nature Astronomy},
         year = 2019,
        month = jan,
       volume = {3},
        pages = {88-92},
          doi = {10.1038/s41550-018-0610-1},
archivePrefix = {arXiv},
       eprint = {1807.04275},
 primaryClass = {astro-ph.HE},
       adsurl = {https://ui.adsabs.harvard.edu/abs/2019NatAs...3...88G}
}

@ARTICLE{Murase2018-blazar-jet,
       author = {{Murase}, Kohta and {Oikonomou}, Foteini and {Petropoulou}, Maria},
        title = "{Blazar Flares as an Origin of High-energy Cosmic Neutrinos?}",
      journal = {\apj},
         year = 2018,
        month = oct,
       volume = {865},
       number = {2},
          eid = {124},
        pages = {124},
          doi = {10.3847/1538-4357/aada00},
archivePrefix = {arXiv},
       eprint = {1807.04748},
 primaryClass = {astro-ph.HE},
       adsurl = {https://ui.adsabs.harvard.edu/abs/2018ApJ...865..124M}
}

@ARTICLE{Inoue2020-AGN-corona,
       author = {{Inoue}, Yoshiyuki and {Khangulyan}, Dmitry and {Doi}, Akihiro},
        title = "{On the Origin of High-energy Neutrinos from NGC 1068: The Role of Nonthermal Coronal Activity}",
      journal = {\apjl},
         year = 2020,
        month = mar,
       volume = {891},
       number = {2},
          eid = {L33},
        pages = {L33},
          doi = {10.3847/2041-8213/ab7661},
archivePrefix = {arXiv},
       eprint = {1909.02239},
 primaryClass = {astro-ph.HE},
       adsurl = {https://ui.adsabs.harvard.edu/abs/2020ApJ...891L..33I}
}
\bibliographystyle{aasjournal}

\end{document}